\documentclass[a4paper,11pt]{article}

\renewcommand{\thetable}{\Roman{table}}

\pdfoutput=1 % if your are submitting a pdflatex (i.e. if you have
\usepackage{jheppub} % for details on the use of the package, please
\usepackage[T1]{fontenc} % if needed
\usepackage{lineno,hyperref}
\usepackage{caption}
\usepackage{subcaption}
\usepackage{soul}
\title{\boldmath Enhancing resonant circular-section haloscopes for dark matter axion detection: approaches and limitations in volume expansion}

\author[*,a]{J.M.~Garc\'ia-Barcel\'o,}
\author[b]{A.~D\'iaz-Morcillo,}
\author[c]{B.~Gimeno}

\affiliation[a]{Max-Planck-Institut f\"{u}r Physik (Werner-Heisenberg-Institut), F\"{o}hringer Ring 6, 80805 M\"{u}nchen, Germany}
\affiliation[b]{Department of Information Technologies and Communications, Universidad Politécnica de Cartagena, 30203 - Cartagena, Spain}
\affiliation[c]{Instituto de F\'isica Corpuscular (IFIC), CSIC-University of Valencia, 46980 - Valencia, Spain}

\affiliation[*]{Corresponding author}

\emailAdd{jmgarcia@mpp.mpg.de}

\abstract
{Haloscopes, microwave resonant cavities utilized in detecting dark matter axions within powerful static magnetic fields, are pivotal in modern astrophysical research. This paper delves into the realm of cylindrical geometries, investigating techniques to augment volume and enhance compatibility with dipole or solenoid magnets. The study explores volume constraints in two categories of haloscope designs: those reliant on single cavities and those employing multicavities. In both categories, strategies to increase the expanse of elongated structures are elucidated. For multicavities, the optimization of space within magnets is explored through 1D configurations. Three subcavity stacking approaches are investigated, while the foray into 2D and 3D geometries lays the groundwork for future topological developments. The results underscore the efficacy of these methods, revealing substantial room for progress in cylindrical haloscope design. Notably, an elongated single cavity design attains a three-order magnitude increase in volume compared to a WC-109 standard waveguide-based single cavity. Diverse prototypes featuring single cavities, 1D, 2D, and 3D multicavities highlight the feasibility of leveraging these geometries to magnify the volume of tangible haloscope implementations.}

\begin{document}
\maketitle
\flushbottom

\section{Introduction}
\label{s:Introduction}

In recent decades, there has been a lot of interest in axions and other particles compatible with the Standard Model that could be part of dark matter. Axions, the particles proposed by Weinberg \cite{Weinberg:1978} and Wilczek \cite{Wilczek:1978}, might address the strong Charge Conjugation-Parity problem \cite{Peccei:1977Jun,Peccei:1977Sep}. A few years later, utilising the misalignment hypothesis, it was projected that axions may perhaps constitute dark matter \cite{Preskill:1983,Abbott:1983,Dine:1983}.\\

Several research groups have constructed during the last thirty years experimental devices to look for axions, which are based on the inverse Primakoff effect \cite{Primakoff:1951,Irastorza:2018dyq}. These detection methods are further classified into three varieties based on the origin of the axion source: haloscopes, helioscopes, and Light Shining through Walls (LSW). The last one produces axion particles intentionally in the laboratory. However, haloscopes and helioscopes rely on external natural sources (relic axions from the galactic halo or axions from the Sun, respectively). They are all based on the axion-photon conversion, which is boosted by an external magnet with a high level of magnetostatic field. Furthermore, in the case of haloscopes, this conversion can be enhanced by using high-quality factor resonators (such as microwave cavities).\\

At present, the forefront of experiments aimed at detecting axions is centered on the haloscope framework, remarkably exemplified by the ADMX (Axion Dark Matter eXperiment) \cite{Stern:2016} and CAPP (Center for Axion and Precision Physics) \cite{Andrew:2023} initiatives. These groups operate below $1$~GHz and above $1$~GHz frequency domains, respectively. In the helioscope scenario, the CAST (CERN Axion Solar Telescope) investigation \cite{Anastassopoulos:2017ftL}, albeit decommissioned, and the forthcoming IAXO (International Axion Observatory) \cite{Armengaud_2014} project assume fundamental roles.\\

In anticipation of IAXO, the BabyIAXO magnet prototype emerges as a first prototype, affording a versatile platform for both helioscope and haloscope axion experiments \cite{BabyIAXO}. The high-volume magnetic configurations featured in the IAXO and BabyIAXO magnets not only demonstrate cost-effectiveness at low frequencies \cite{RADES_BabyIAXO_ArXiv} but also facilitate the establishment of haloscopes with large volumes operating at higher frequencies \cite{Volume_paper}. These experimental efforts are actively working to achieve sensitivities aligned with the KSVZ (Kim-Shifman-Vainshtein-Zakharov) and DFSZ (Dine-Fischler-Srednicki-Zhitnitsky) models, which offer promising potential for detecting axions \cite{BabyIAXO}.\\

Remarkably, the ADMX collaboration has achieved significant sensitivities within the $2.81-3.31$~$\mu$eV range \cite{Braine:2019fqb}. Noteworthy is the recent proposition by the CAPP consortium, outlining an axion dark matter exploration through a DFSZ-sensitive haloscope spanning the $4.51$ to $4.59$~$\mu$eV axion mass range \cite{Andrew:2023}. Furthermore, the HAYSTAC (Haloscope At Yale Sensitive To Axion Cold Dark Matter) group has achieved KSVZ-sensitive outcomes from dual independent searches for dark matter axions within the $16.96$ to $17.28$~$\mu$eV and $23.15$ to $24.0$~$\mu$eV axion mass intervals \cite{Backes:2020ajv,Zhong:2018}. Clearly, these experimental teams are making progress in achieving sensitivities that hold significant theoretical importance within the realm of haloscope investigations.\\

Beyond the aforementioned initiatives, other axion groups like RADES \cite{RADES_paper3}, QUAX \cite{Alesini:2022}, and FLASH \cite{Alesini:2017ifp,Alesini:2023qed} have also made noticeable progress in enhancing the utilization of haloscope methodologies in recent times.\\

The whole setup of the haloscope involves a mix of essential parts. First of all, a suitable cavity designed to operate at the frequency where the axion search will take place is required. On the other hand, due to the extremely weak interaction between axions and photons, it's crucial to maintain a cryogenic environment at temperatures in the Kelvin range. This helps minimize any unwanted heat effects. In addition, for boosting the axion-photon conversion, a high static magnetic field is needed. Thus, a magnet is necessary, where the resonant cavity will be installed. After that, the radio frequency (RF) power collected within the haloscope goes through a series of changes including amplification, filtering, and down-conversion, culminating in its interception by a receiver.\\

At the final stage, the receiver steps in to handle main operations, including converting analog signals into digital format. This is followed by a process of data manipulation, achieved using Fast Fourier Transform algorithms, a standard technique widely implemented in the theory of signal processing \cite{RADES_paper3}.\\

A fundamental objective inherent in the design of an axion detection configuration revolves around the augmentation of the axion-photon conversion sensitivity in the haloscope device. This increase, which depends on a specific signal-to-noise ratio ($\frac{S}{N}$), can be easily formulated through \cite{RADESreviewUniverse}
\begin{equation}
\label{eq:ga}
    g_{a\gamma} \, = \,  \left(\frac{\frac{S}{N} \, k_B \, T_{sys} \, \left(1+\beta\right)^2}{\rho_a \, C \, V \, \beta \, Q_0}\right)^{\frac{1}{2}}\frac{1}{B_e}\left(\frac{m_a^3}{Q_a \, \Delta t}\right)^{\frac{1}{4}}, \text{provided that } Q_a \gg \frac{Q_0}{1+\beta}
\end{equation}
where $\frac{S}{N}$ is the signal-to-noise ratio, $k_B$ is the Boltzmann constant, $T_{sys}$ is the noise temperature of the system expressed in K, $\beta$ is the extraction coupling factor, $\rho_a$ is the dark matter volumetric density, $C$ is the form factor of the cavity mode, $V$ is the volume of the haloscope, $Q_0$ is the unloaded quality factor, $B_e$ is the maximum intensity of the external magnetostatic field $\vec{B}_e$, $m_a$ is the axion mass (directly proportional to the operational frequency of the experiment), $Q_a$ is the axion quality factor, and $\Delta t$ is the time window used in the axion data campaign. A requisite condition for achieving optimal power transfer is the attainment of a $\beta$ value equal to unity ($\beta=1$), the so-named critical coupling regime. It is crucial to emphasize that the properties of the external static magnetic field ($\vec{B}_e$) depend on the particular magnet arrangement utilized in the experimental setup, typically being of either solenoidal or dipolar type. Furthermore, the spatial distribution and polarization of this magnetic field need to be analysed carefully since they play a key role in selecting the operation mode (axion mode) in the resonant device.\\

Despite its pronounced high-Q resonant characteristics, the ultimate objective of a cavity-based haloscope is the exploration of a wide range of mass values accomplished in a time-efficient manner. The process of mass variation encompassing a breadth of values is facilitated through the integration of efficient mechanical mechanisms or the strategic deployment of materials capable of shifting the resonant frequency of the cavity. The quantification of the rate at which this mass exploration runs is denoted as the scanning rate, often employed as a key parameter encapsulating the efficacy and performance of the haloscope.\\

In the context of conducting a measurement wherein the resonance of the cavity is aligned with the axion mass $m_a$, considering the temporal requisites for effecting achieve a predetermined level of $\frac{S}{N}$, the scanning rate can be expressed as \cite{RADESreviewUniverse}
\begin{equation}
\label{eq:dmadt}
    \frac{dm_a}{dt} = Q_aQ_0\frac{\beta^2}{\left(1+\beta\right)^3} \, g_{a\gamma}^4\left(\frac{\rho_a}{m_a}\right)^2B_e^4C^2V^2\left(\frac{S}{N}k_BT_{sys}\right)^{-2}.
\end{equation}

The form factor serves as a parameter that gives the parallelism between the external static magnetic field ($\vec{B}_e$) and the radio frequency electric field ($\vec{E}$), the latter being induced within the cavity through the axion-photon interaction. This key indicator is mathematically characterized as follows:
\begin{equation}
\label{eq:C}
    C \, = \, \frac{|\int _V \, \vec{E} \cdot \vec{B}_e \, dV|^2}{\int_V \, |\vec{B}_e|^2 \, dV \int_V \, \varepsilon_r \, |\vec{E}|^2 \, dV}.
\end{equation}
In this equation, the symbol $\varepsilon_r$ stands representative of the relative electric permittivity present in the cavity medium, which predominantly comprises air or vacuum. Thus, the modifiable and optimizable facets within the design of a haloscope resonator encompass this set of parameters: $\beta$, $C$, $V$, and $Q_0$. In this work, in order to measure the performance of the cavity structure, we will use $Q_0V^2C^2$, following \ref{eq:dmadt}.\\

The principal objective of this investigation revolves around an analysis for amplifying the volumetric dimensions of haloscopes employing cylindrical structures. As a result of this expansion, enhanced sensitivity and scanning rate for the detection of axions are envisaged. Noteworthy precedent explorations pertaining to the viability of rectangular cavities were previously conducted and discussed in \cite{Volume_paper}. In the present article, a similar study is intended to be carried out for haloscopes with a circular cross-section, which is more common in dark matter axion experiments, since it makes better use of the volume of the cylindrical bores of the magnets. For a haloscope design, the maximum admissible volume is essentially determined by four key factors that interact together: the geometry of the cavity, the specific electromagnetic mode and the operational frequency employed, the use of multicavity approaches, and the magnet system's geometry and typology (how exactly the magnetic field is aligned for measuring axions).\\

The characteristic resonant frequency associated with cylindrical cavities engaged in $TE_{mnp}$ and $TM_{mnp}$ modes can be expressed as follows:
\begin{equation}\label{eq:frmnl_cyl_TE}
    f_{TE_{mnp}} = \frac{c}{2\pi \, \sqrt{\varepsilon_r \, \mu_r}}\sqrt{\left(\frac{\chi^{'}_{mn}}{a}\right)^2+\left(\frac{p \, \pi}{d}\right)^2},
    \begin{alignedat}{2}
    \qquad m=0,1,2,...\\
    \qquad n=1,2,3,...\\
    \qquad p=1,2,3,...\\
    \end{alignedat}
\end{equation}
\begin{equation}\label{eq:frmnl_cyl_TM}
    f_{TM_{mnp}} = \frac{c}{2\pi \, \sqrt{\varepsilon_r \, \mu_r}}\sqrt{\left(\frac{\chi_{mn}}{a}\right)^2+\left(\frac{p \, \pi}{d}\right)^2},
    \begin{alignedat}{2}
    \qquad m=0,1,2,...\\
    \qquad n=1,2,3,...\\
    \qquad p=0,1,2,...\\
    \end{alignedat}
\end{equation}
where $c$ is the speed of light in vacuum; $\mu_r$ is the relative magnetic permeability ($\mu_r=\varepsilon_r=1$ is imposed in this paper); $m$, $n$, and $p$ are integers that denote the number of maxima of the electric field magnitude along the $\varphi$ (azimut), $\rho$ (radius), and $z$ (length) axes, respectively; $\chi_{mn}$ and $\chi^{'}_{mn}$ are the $n-$th zeros of the Bessel function $J_{m}$ and its derivate $J^{'}_{m}$ of order $m$, respectively; and $a$ and $d$ are the radii and length of the cavity, respectively. As highlighted by the equation provided, the frequencies at which resonance occurs depend on the size of the cavity. This mathematical connection shows the difficulties of trying to increase the volume of the cavity and the operational frequency at the same time. In addition, the mode clustering should be avoided, which could introduce higher complexity in the cavity design pointing out the potential challenges in achieving these two goals simultaneously.\\

The challenge of mode clustering, dictating the interaction between the principal mode and its subsequent higher order counterparts, can result in a diminution in the form factor parameter of the desired mode. This issue persists even when higher order modes manifest form factors attaining a null value.\\

\begin{figure}[h]
\centering
\begin{subfigure}[b]{0.35\textwidth}
         \centering
         \includegraphics[width=1\textwidth]{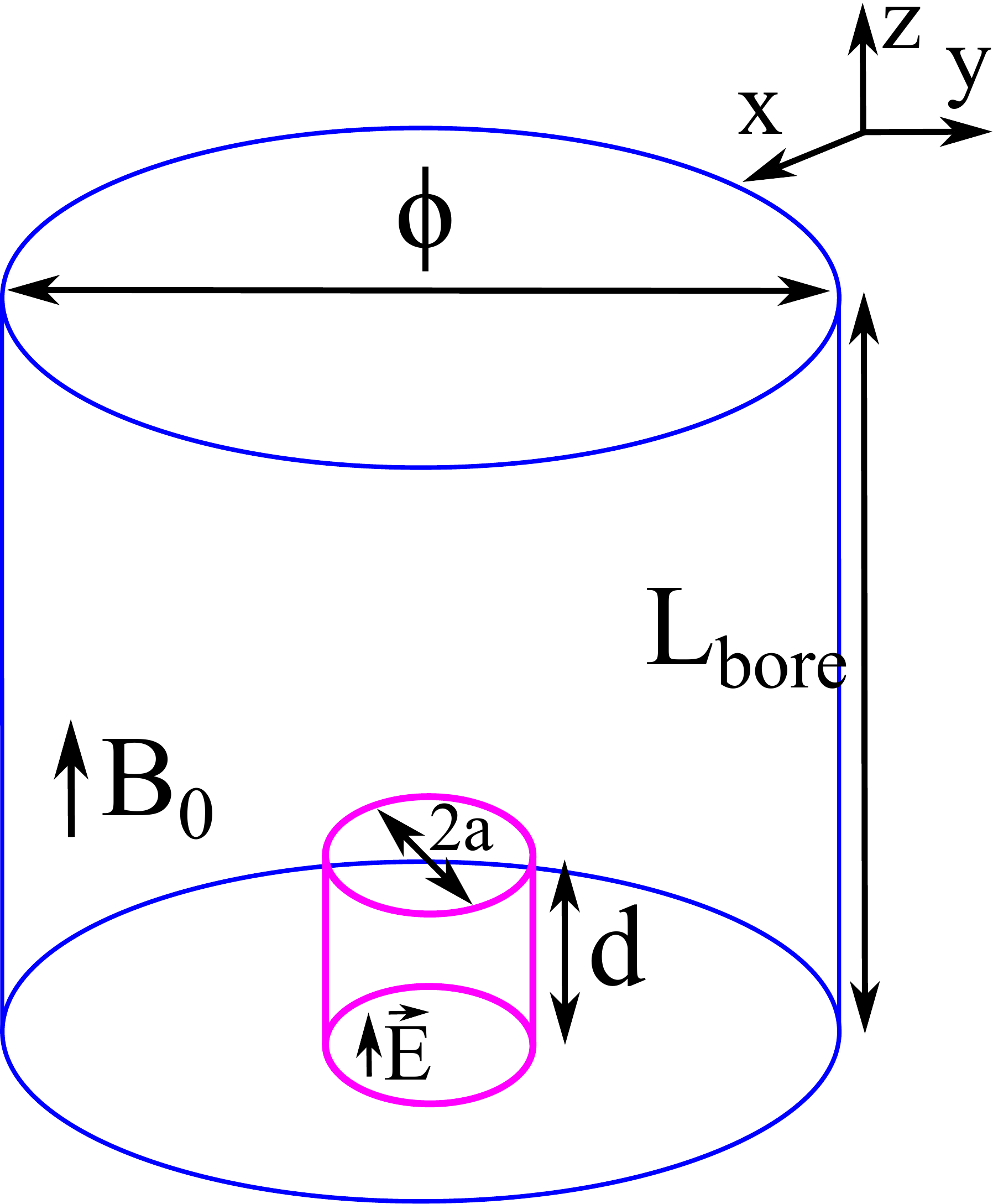}
         \caption{}
         \label{fig:SingleCavity_Solenoid}
\end{subfigure}
\hfill
\begin{subfigure}[b]{0.6\textwidth}
         \centering
         \includegraphics[width=1\textwidth]{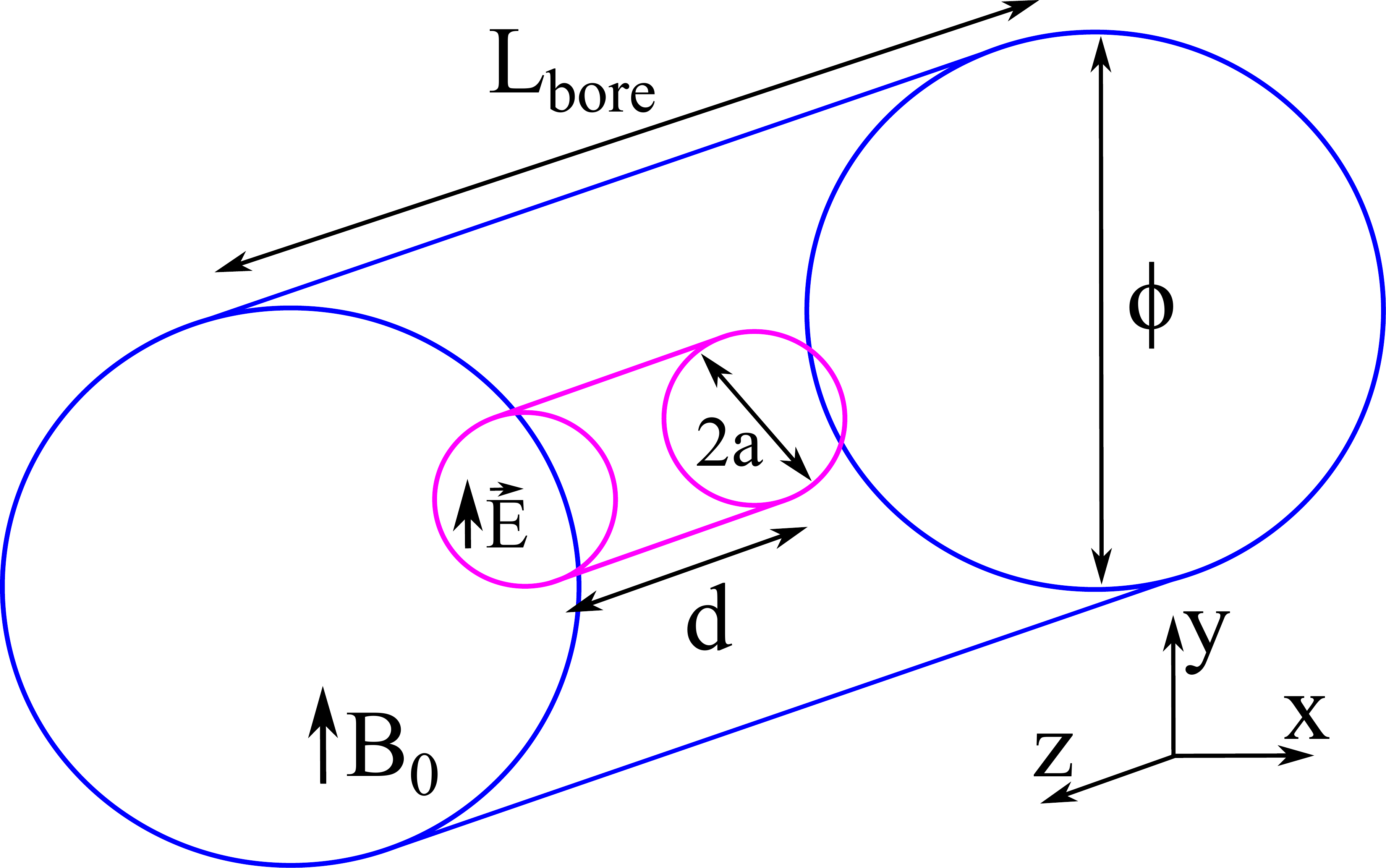}
         \caption{}
         \label{fig:SingleCavity_Dipole}
\end{subfigure}
\caption{(a) Solenoid magnet (blue) featuring a cylindrical cavity (pink) characterized by a radius $a$ and a length $d$, designed to operate within the ambit of the $TM_{010}$ cylindrical mode, and (b) dipole magnet (blue) housing a similar cavity (pink) but now working with the $TE_{111}$ cylindrical mode. Notably, for the solenoid magnet, the predominant orientation of the magnetic field primarily aligns with the $z-$axis, whereas for the dipole magnet this alignment is predominantly directed along the $y-$axis. It is crucial to remark that although the illustration does not overtly indicate, it is assumed that the probes designated for external couplings shall be situated along the upper boundaries of the cavities, aligning with the $z-$axis in (a) and the $y-$axis in (b).}
\label{fig:SingleCavity_Solenoid&Dipole}
\end{figure}

Solenoids, as depicted in Figure~\ref{fig:SingleCavity_Solenoid}, constitute the predominant selection of magnets deployed within experimental frameworks for the detection of dark matter axions.
These solenoidal magnets have a specialized role in important research projects, as demonstrated by their use in notable initiatives like ADMX \cite{Braine:2019fqb}, CAPP \cite{Andrew:2023}, and HAYSTAC \cite{PhDThesis-Brubaker}. The axial ($z-$axis) static magnetic field they generate is aligned with the electric field of the $TM_{010}$ cylindrical mode, contributing to an optimal form factor. In contrast, accelerator dipole magnets (see Figure~\ref{fig:SingleCavity_Dipole}), as the CAST magnet, give a robust transverse static magnetic field, which necessitates the employment of a cylindrical cavity operating under the $TE_{111}$ mode. This setup mainly leads to vertical polarization, where the RF electric field aligns mostly parallel to the static magnetic field in the dipole arrangement \cite{RADES_BabyIAXO_ArXiv,Adair:2022}.\\

Furthermore, an illustrative example worth considering is the future BabyIAXO experiment, which deploys a superconducting toroidal magnet, and its magnetic field pattern is presented in \cite{BabyIAXO}. Nevertheless, for the purpose of this paper, the BabyIAXO configuration can be conveniently treated as akin to a dipole magnet. It is pertinent to note that in this study, for the purpose of calculating the form factor, a simplified approximation implies the assumption of a uniform static magnetic field $\vec{B}_e=B_e \, \hat{y}$ for the dipole and quasi-dipole magnet (i.e. BabyIAXO) configurations \cite{BabyIAXO}.\\

A comprehensive overview encompassing the most commonly utilized magnets by several research groups is shown in \cite{Volume_paper}. For a clarification of the optimal alignment of cylindrical cavities with solenoid and dipole magnets, refer to Figure~\ref{fig:SingleCavity_Solenoid&Dipole}.\\

As a general trend, the aspect ratio characterizing the bore of dipole magnets leads towards a considerably large configuration relative to their diameter. Conversely, the bores of solenoid magnets usually have dimensions where the length and diameter look more alike. Consequently, the main objective for increasing the volume in the haloscope involves calibrating the cavity geometry to effectively exploit the available room within the magnet bore. One of the pioneering efforts that aimed to make the most of elongated cavities was seen in the development and execution of a toroidal-shaped cavity, an initiative from the CAPP team \cite{Ko:2016,Choi:2017}. Notably, in this scenario, the BabyIAXO magnet steps up as a perfect fit for this pursuit. Nonetheless, it's worth noting the rise of novel geometries as potential paths to explore, like incorporating large structures in rectangular setups, which presents an attractive idea \cite{Volume_paper}.\\

It is important to point out here other interesting concepts for obtain high frequencies-high volumes, not related to multicavities, as wire array \cite{Wooten:2023} or dielectric array \cite{Bae:2023} metamaterial cavities, or phase-matched multiple cavities \cite{Adair:2022}.\\

In Section~\ref{s:SingleCavities}, we expose the thorough investigations conducted aiming to uncover the limitations of volumetric capacities in haloscopes for long single cylindrical cavities operating with $TM_{010}$ and $TE_{111}$ modes. The central objective is to establish these critical thresholds, and a meticulous analysis is dedicated to scrutinizing frequency spacing (mode clustering) examining the form factor, quality factor and volume parameters for the optimization of axion detection.\\

Following this, Section~\ref{s:1Dmulticavities} introduces a new concept: one-dimensional (1D) multicavity configurations based on cylindrical subcavities interconnected via iris windows. Three spatial orientations are analysed for the couplings between subcavities: $\varphi$, $\rho$, and $z$ axes. In addition, two scenarios are explored in these haloscopes: dipole and solenoid configurations, considering the magnets presented in \cite{Volume_paper}. To see how these new setups measure up against standard single cavities, a detailed comparison by examples that work in the X-band frequency range ($8-12$~GHz) is shown, analysing the key cavity parameters (form factor, quality factor, and volume). Furthermore, this paper systematically clarifies approaches aimed at enhancing the volume of cylindrical subcavities, encompassing both length and diameter.\\

Section~\ref{s:2D3Dmulticavities} establishes a framework based on the expansion of the multicavity concept into the realms of two-dimensional (2D) and three-dimensional (3D) structures. In this study, the combination of different stacking orientations for the coupling between subcavities is explored. The discussion introduces various illustrative examples of 2D and 3D multicavity arrangements. These instances serve as effective means to clarify the benefits provided by these evolved structures when compared to their one-dimensional (1D) counterparts highlighting its potential in optimizing the efficient utilization of magnets for axion search experiments.\\

Finally, Section~\ref{s:Conclusions} shows the conclusions of the present investigations detailing future lines of work arising from these studies.

\section{Single cavities}
\label{s:SingleCavities}

The $TM_{010}$ mode is used for cylindrical cavities operating in solenoid magnets because it maximises the form factor stated in equation~\ref{eq:C}. Because the length of the cavity $d$ has no effect on the resonant frequency in this mode, it can be adjusted as desired to expand the cavity volume. However, there is a limit where $d$ cannot be raised owing to the proximity of higher order modes with $p\neq0$ (in this case, the nearest mode is the $TM_{011}$). This scenario may impede accurate mode identification and, in some situations, may lower the form factor.\\

On the other hand, in dipole magnets the $TE_{111}$ is usually employed working with cylindrical cavities. The simplest way to raise the length of the haloscope without decreasing the resonant frequency is to slightly reduce the radius $a$, according to equation~\ref{eq:frmnl_cyl_TE}. Because this reduction is little in comparison to the length gained, the overall volume will be increased. Furthermore, when the cavity length is substantially greater than its radius, the resonant frequency becomes virtually independent of the cavity length ($f_{TE_{111}}\approx\frac{c\chi_{11}^{'}}{2\pi a}$). Again, the restriction is enforced by the closeness of the next resonant mode (in this case, with the $TE_{112}$ mode).

\subsection{Long cavities}
\label{ss:SingleCavities_Long}

\begin{figure}[h]
\centering
\includegraphics[width=0.55\textwidth]{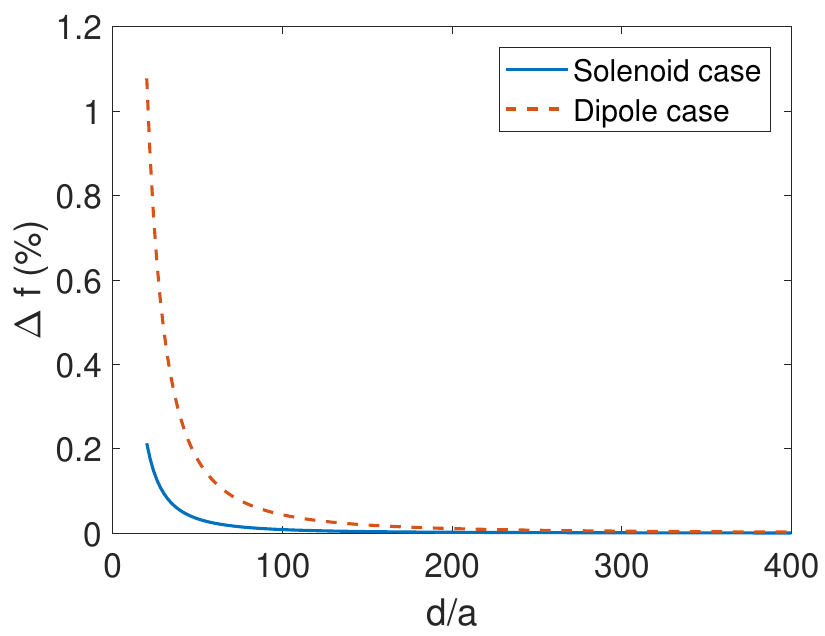}
\caption{Relative mode separation between modes $TM_{010}$ and $TM_{011}$ (solenoid case) (blue solid curve) and between modes $TE_{111}$ and $TE_{112}$ (dipole case) (red dashed curve) versus $d/a$.}
\label{fig:ModeClustering_vs_d-a_Solenoid_vs_Dipole_pdf}
\end{figure}

Figure~\ref{fig:ModeClustering_vs_d-a_Solenoid_vs_Dipole_pdf} shows the relative mode separation $\Delta f$ for a cylindrical cavity as a function of $d/a$ for both scenarios, solenoid and dipole.
\\

This mode separation is defined as
\begin{equation}\label{eq:Deltaf}
    \Delta f=\frac{|f_{axion}-f_{neighbour}|}{f_{axion}}\times 100 \quad [\%]
\end{equation}
where $f_{axion}$ is the resonant frequency of the mode excited by the axion-photon conversion, and $f_{neighbour}$ is the resonant frequency of the nearest mode.\\

The findings demonstrate a rapid decline in mode separation as the parameter $d/a$ increases. When the subsequent mode is sufficiently distant in frequency, the form factor reaches its theoretical maximum value for any cavity size, namely $C_{TM_{010}}=0.6917$ and $C_{TE_{111}}=0.6783$, derived from equation~\ref{eq:C}.\\

\begin{figure}[h]
\centering
\begin{subfigure}[b]{0.49\textwidth}
         \centering
         \includegraphics[width=1\textwidth]{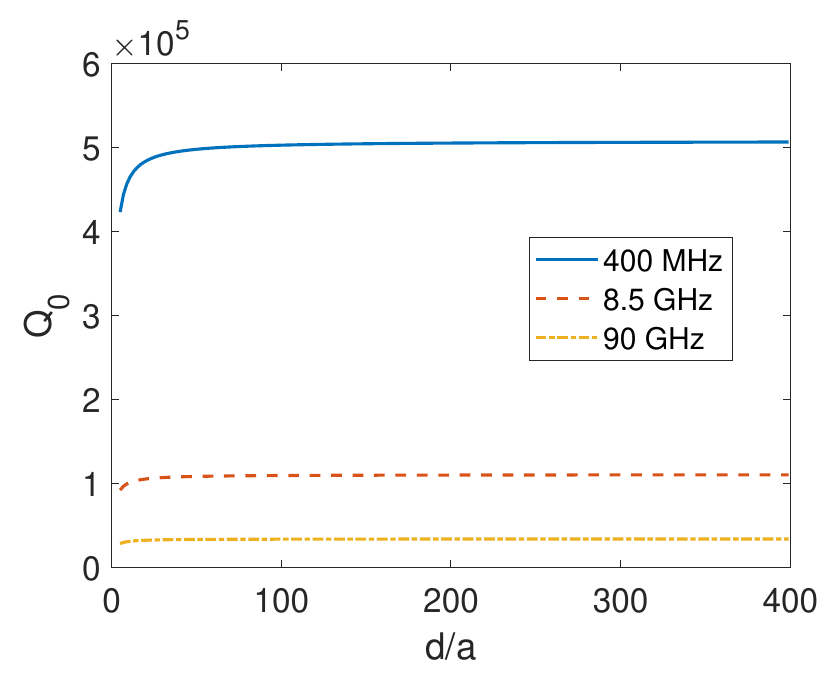}
         \caption{}
         \label{fig:Q0_TMmnl_vs_d-a_Bands}
\end{subfigure}
\hfill
\begin{subfigure}[b]{0.49\textwidth}
         \centering
         \includegraphics[width=1\textwidth]{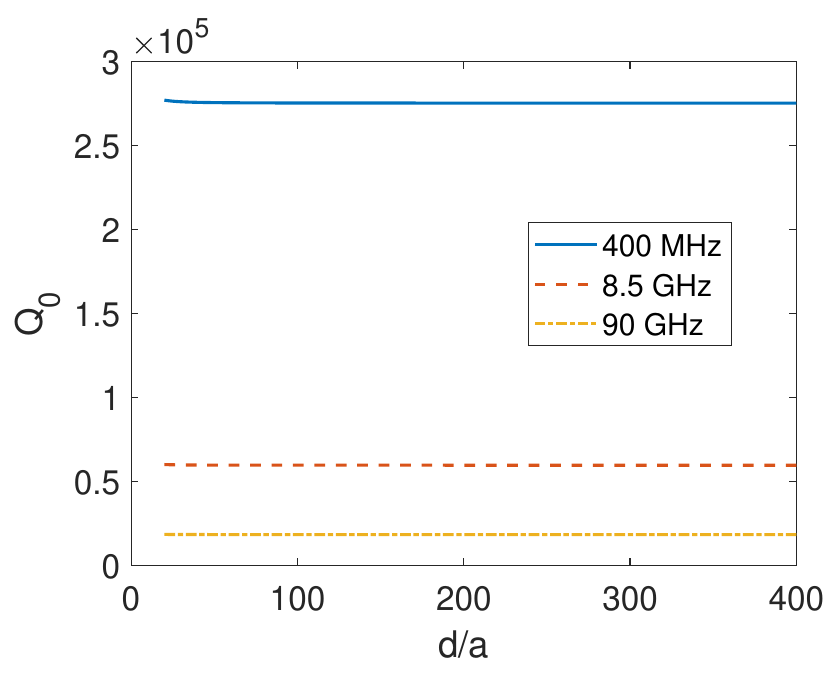}
         \caption{}
         \label{fig:Q0_TEmnl_vs_d-a_Bands}
\end{subfigure}
\caption{(a) $Q_0$ of the $TM_{mnl}$ and (b) $TE_{mnl}$ modes as a function of $d/a$ for three frequencies ($0.4$, $8.4$, and $90$~GHz).}
\label{fig:Q0_vs_d-a_Bands}
\end{figure}

For a cylindrical cavity, the quality factors of the $TM_{mnl}$ and $TE_{mnl}$ modes can be expressed as \cite{collin_FMI}
\begin{equation}\label{eq:Q0_TMmnl}
Q_{TM_{mnl}}=\left\{ 
  \begin{array}{ c l }
    \frac{\lambda_0}{\delta_c}\frac{\left[\chi_{mn}^2 + (l\pi a/d)^2\right]^{1/2}}{2\pi(1+a/d)}, & \quad l>0 \\\\
    \frac{\lambda_0}{\delta_c}\frac{\chi_{mn}}{2\pi(1+a/d)},                 & \quad l=0
  \end{array}
\right.
\end{equation}
\begin{equation}\label{eq:Q0_TEmnl}
Q_{TE_{mnl}}=\frac{\lambda_0}{\delta_c}\frac{\left[1 - \left(\frac{m}{\chi^{'}_{mn}}\right)^2\right]\left[(\chi^{'}_{mn})^2 + \left(\frac{l\pi a}{d}\right)^2\right]^{3/2}}{2\pi\left[(\chi^{'}_{mn})^2 + \frac{2a}{d}\left(\frac{l\pi a}{d}\right)^2 + \left(1 - \frac{2a}{d}\right)\left(\frac{ml\pi a}{\chi^{'}_{mn}d}\right)^2\right]}
\end{equation}
where $\lambda_0=c/f_r$ is the free-space wavelength of the mode, $f_r$ is the resonant frequency of the cavity mode, $\delta_c = 1/(\sqrt{\pi\mu_0\sigma_c f_r})$ is the skin depth of the cavity surface, $\mu_0$ is the vacuum magnetic permeability, and $\sigma_c$ is the electrical conductivity of the cavity walls (a conductivity value of $\sigma_c=2\times 10^9$~S/m is adopted, aligning with the characteristics of copper at cryogenic temperatures). By plotting the aforementioned equations (refer to Figures~\ref{fig:Q0_TMmnl_vs_d-a_Bands} and \ref{fig:Q0_TEmnl_vs_d-a_Bands}), it becomes evident that the unloaded quality factor experiences a decline at elevated frequencies in both scenarios, corresponding to a reduction in the cavity radius.
Figure~\ref{fig:Q0_vs_d-a_Bands} allows us to infer that the parameter $Q_0$ remains unaffected by the cavity length for high values of $d$ in both modes.\\

\begin{figure}[h]
\centering
\begin{subfigure}[b]{0.49\textwidth}
         \centering
         \includegraphics[width=1\textwidth]{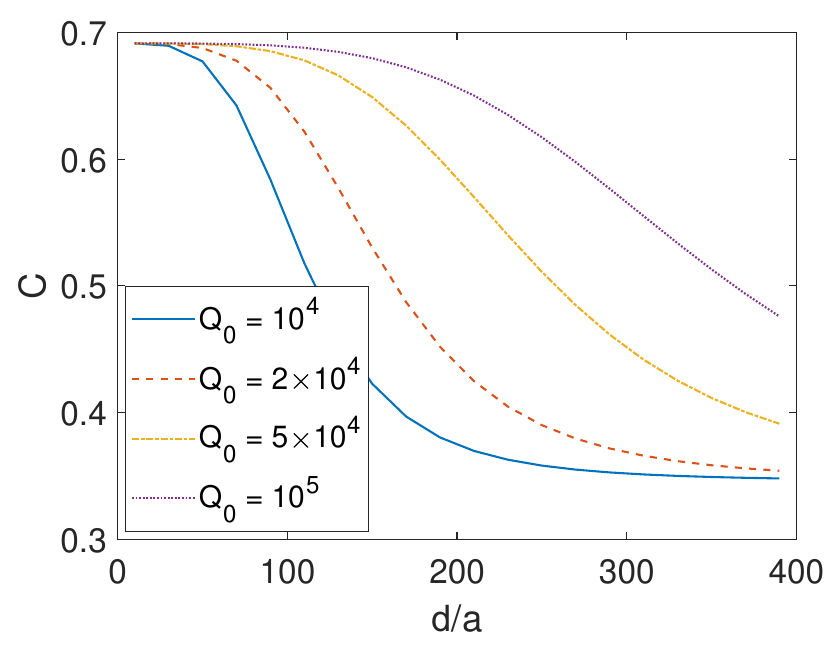}
         \caption{}
         \label{fig:C_vs_d-a_vs_Q0_Solenoid}
\end{subfigure}
\hfill
\begin{subfigure}[b]{0.49\textwidth}
         \centering
         \includegraphics[width=1\textwidth]{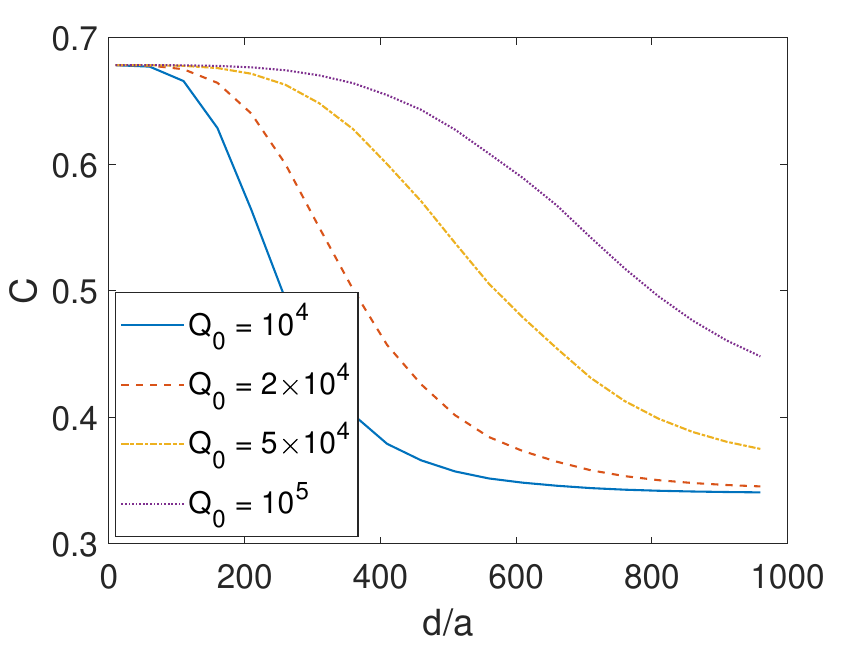}
         \caption{}
         \label{fig:C_vs_d-a_vs_Q0_Dipole}
\end{subfigure}
\caption{Form factor as a function of the $d/a$ parameter for several $Q_0$ values for (a) the $TM_{010}$ mode, and (b) the $TE_{111}$ mode.}
\label{fig:C_vs_d-a_vs_Q0_Solenoid_and_Dipole}
\end{figure}

The minimum accepted mode separation (mode clustering) relies on the measured quality factor, which in turn is influenced by the cavity's shape, material, and manufacturing process quality. Higher $Q_0$ values result in sharper resonances, allowing modes to approach each other in frequency without degradation. Generally, taking a conservative approach, it can be anticipated that the unloaded quality factor of the fabricated prototype will be approximately half of the theoretical value due to manufacturing tolerances in the fabrication process (such as inner wall roughness, soldering quality, and metallic contact in the case of screws usage). To quantify the impact of mode clustering on energy loss, Figures~\ref{fig:C_vs_d-a_vs_Q0_Solenoid} and \ref{fig:C_vs_d-a_vs_Q0_Dipole} illustrate the form factor plotted against $d/a$ for various $Q_0$ values ($Q_0=10^4$ for the blue line, $Q_0=2\times10^4$ for the red line, $Q_0=5\times10^4$ for the yellow line, and $Q_0=10^5$ for the purple line) in the solenoid and dipole scenarios, respectively.
This plot demonstrates that higher $Q_0$ values correspond to a reduced deterioration in $C$. The form factor presented in Figures~\ref{fig:C_vs_d-a_vs_Q0_Solenoid} and \ref{fig:C_vs_d-a_vs_Q0_Dipole} has been computed using equation~\ref{eq:C}, which accounts for the perturbation of the electric field and its resulting impact on $C$ due to the influence of the electric field generated by the adjacent resonant modes ($TM_{011}$ and $TE_{112}$ modes in this case, respectively) when they are in close proximity. The detailed analysis of this phenomenon for rectangular cavities can be found in \cite{Volume_paper}. Consequently, a constraint must be imposed on the mode separation to achieve a minimum form factor.\\

A maximum form factor reduction of $\sim1\%$ is chosen as our least allowable lowered value owing to mode clustering. This gives a form factor value of $C=0.68$ for the $TM_{010}$ mode (solenoid scenario) and $C=0.67$ for the $TE_{111}$ mode (dipole scenario). The form factor could even be more reduced, but very close modes generate problems in the correct measurement of $f_r$ and $Q_0$ in the experiment. In any case, if the cavity response has two resonances that are extremely near or even mixed as a result of lower than expected quality factors, there are ways to recover the original form of each resonance and compute the necessary two parameters ($f_r$ and $Q_0$) \cite{Q0_Characterization}. To extract the maximum dimensions, while maintaining these minimum form factor values, the plots from Figures~\ref{fig:C_vs_d-a_vs_Q0_Solenoid} and \ref{fig:C_vs_d-a_vs_Q0_Dipole} are analysed for the solenoid and dipole case, respectively.\\

This form factor minimum value is attained for the X-band situation with the dimensions provided in Table~\ref{tab:SingleLongQVC}. These figures are based on an unloaded quality factor after manufacture of half of the theoretical one. The advantages gained with these examples in contrast to WC-109 cylindrical cavities are also shown in Table~\ref{tab:SingleLongQVC}.
\begin{table}[h]
\scalebox{0.8}{
\begin{tabular}{|c|c|c|c|c|c|c|c|c|c|}
\hline
Cavity type & Magnet type & $2a$ (mm) & $d$ (mm) & $\Delta f$ ($\%$) & $V$ (mL) & $Q_0$ & $C$ & $Q_0V^2C^2$ (L$^2$) \\ \hline\hline
Standard & Solenoid & $27$ & $26.39$ & $20.27$ & $15.11$ & $7.32\times10^4$ & $0.6917$ & $7.996$ \\ \hline
Standard & Dipole & $27.788$ & $26.39$ & $52.96$ & $16.005$ & $7.89\times10^4$ & $0.6783$ & $9.299$ \\ \hline
Long & Solenoid & $27$ & $1350$ & $0.009$ & $772.95$ & $10.95\times10^4$ & $0.68$ & $3.03\times10^4$ \\ \hline
Long & Dipole & $20.67$ & $1395$ & $0.024$ & $468.107$ & $5.97\times10^4$ & $0.67$ & $5872$ \\ \hline
\end{tabular}
}
\centering
\caption{\label{tab:SingleLongQVC} Comparison of the operational parameters of standard WC-109 cylindrical cavities employed for resonating at $8.5$~GHz at both solenoid and dipole scenarios, with very long cavities for the same resonant frequency.}
\end{table}
The $Q_0V^2C^2$ factors of $3789$ and $632$ for the solenoid and dipoles situations, respectively, are significant improvements over the standard WC-109 cavities.\\

The previous study is an exploration of the dimension limits for a minimal form factor in cylindrical cavities. However, more conservative limits are imposed for the fitting of the cavities inside the magnets taking into account the higher magnet Figure of Merit $FoM = (B_e^2V)/T$ values. For solenoids, the best case is the MRI-EFR magnet, with $B_{eMRI} = 9$~T, $T_{MRI} = 0.1$~K, $\phi_{MRI} = 800$~mm, and $L_{MRI} = 0.513$~m, giving a $FoM_{MRI} = 215.03$~$\frac{T^2m^3}{K}$. For dipoles, the best case will be the BabyIAXO magnet, with $B_{eBI} = 2.5$~T, $T_{BI} = 4.2$~K, $\phi_{BI} = 600$~mm, and $L_{BI} = 10$~m, giving a $FoM_{BI} = 4.207$~$\frac{T^2m^3}{K}$.\\

For the case of a resonant cavity operating with the $TE_{111}$ mode (dipole case), it could fit perfectly into the BabyIAXO magnet using the boundary dimensions shown in Table~\ref{tab:SingleLongQVC} ($d = 1.4$~m versus $L_{BI} = 10$~m). However, for the case of a cavity operating with $TM_{010}$ resonant mode (solenoid case), it would have to use a shorter length than shown in Table~\ref{tab:SingleLongQVC} to fit the MRI magnet ($d = L_{MRI} = 513$~mm), giving the following parameter values: $\Delta f = 0.059$~$\%$, $V=293.72$~mL, $Q_0=1.08\times10^5$, $C=0.69$, and $Q_0V^2C^2=4424$~L$^2$, which also represent substantial improvements over the standard cavity case.

\subsection{Tuning}
\label{ss:SingleCavities_Tuning}

Various tuning mechanisms are at the disposal for potential adaptation to cylindrical single cavities operating in both solenoid and dipole magnets. For a solenoid scenario, noteworthy among these tuning systems are the mechanical tuning techniques developed by the ADMX group, wherein the manipulation of angular rotations of metal rods serves as a frequency tuning mechanism. This approach has manifested significant outcomes across diverse axion search experiments \cite{Braine:2019fqb}. The HAYSTAC experimental group has similarly implemented the application of a tuning method based on metallic rods \cite{Zhong:2018}. In addition, the CAPP group has utilized a tuning configuration encompassing the deployment of a single dielectric rod positioned within a cavity. This innovative tuning technique is further enhanced by including a piezoelectric rotational actuator, including the controlled circular movement of the rod installed within the cavity \cite{Jeong:2018}.\\

On the other hand, for dipole magnets, the RADES group has developed a series of experiments based on the exploration of rotary mechanisms employing metallic plates. This investigatory trajectory has been accomplished in the context of long single cavities operating at UHF frequencies, dealing with the mode clustering parameter issue \cite{RADES_BabyIAXO_ArXiv}. In addition, the CAPP-CAST group introduced in \cite{Adair:2022} a mechanical tuning system based on two sapphire strips positioned symmetrically along the longitudinal edges. These strips are adjusted simultaneously, moving towards the centre of the cavity.\\

The investigation of tuning techniques for cylindrical cavities with large dimensions has been left as future work. This study is envisaged to be developed by the adaptation of the mechanical and electronic methods evidenced in \cite{3CavityRADES:2019} and \cite{Garcia:2023}, respectively. It is important to mention that by introducing tuning in haloscopes, mode clustering will most likely produce significant mode crossings.

\section{1D multicavities}
\label{s:1Dmulticavities}

The utilization of the multicavity principle in cylindrical haloscope configurations results in a noticeable increase in volume across various spatial orientations including $\varphi$, $\rho$, and $z$. Importantly, this expansion occurs while maintaining the resonant frequency, a characteristic similar to what has been observed in rectangular multicavity setups \cite{Volume_paper,RADESreviewUniverse}. In contrast to the concept of elongated cavities used in dipole magnet arrangements, the multicavity designs oriented along the $z-$axis exhibit the capability to accommodate larger-diameter cylindrical waveguides. An example of this is the standard WC-109 for X-band frequencies. Additionally, in the CAPP team, an efficient high-frequency haloscope that adopts a multi-cell cylindrical cavity approach was developed and put into operation \cite{Jeong:2020cwz,Jeong:2023}. This strategy reinforces the volumetric expansion of the haloscopes while preserving the operating frequency.\\

On the other hand, the RADES team has carried out the development and construction of a series of compact haloscope prototypes with rectangular geometries. Within this effort, versions include an all-inductive configuration that combines five subcavities, as well as two alternating designs featuring different quantities of subcavities ($N=6$ and $N=30$, where $N$ represents the number of subcavities). A thorough analysis of these prototypes is presented in a range of works \cite{RADES_paper1,RADES_paper2,RADES_paper3,RADESreviewUniverse,3CavityRADES:2019}.\\

The Ref. \cite{Volume_paper} presents an exploration of high-volume structures with rectangular geometries for single cavities and multicavity structures. Similarly, in this paper cylindrical subcavities are analysed taking into account the phenomenon of mode clustering that emerges as the number of subcavities or volume increases \cite{Volume_paper}. A detailed examination of the three spatial orientations $-$ $\varphi$, $\rho$, and $z$ $-$ for the stacking of cylindrical subcavities to create a multicavity structure has been explored. For each stacking orientation, exemplary designs are discussed presenting the viability of each scenario. Furthermore, an exploration of multicavities with large subcavities is exposed.

\subsection{Stacking in $\varphi$}
\label{ss:1Dmulticavities_StackinginPhi}

\begin{figure}[h]
\centering
\begin{subfigure}[b]{0.3\textwidth}
         \centering
         \includegraphics[width=1\textwidth]{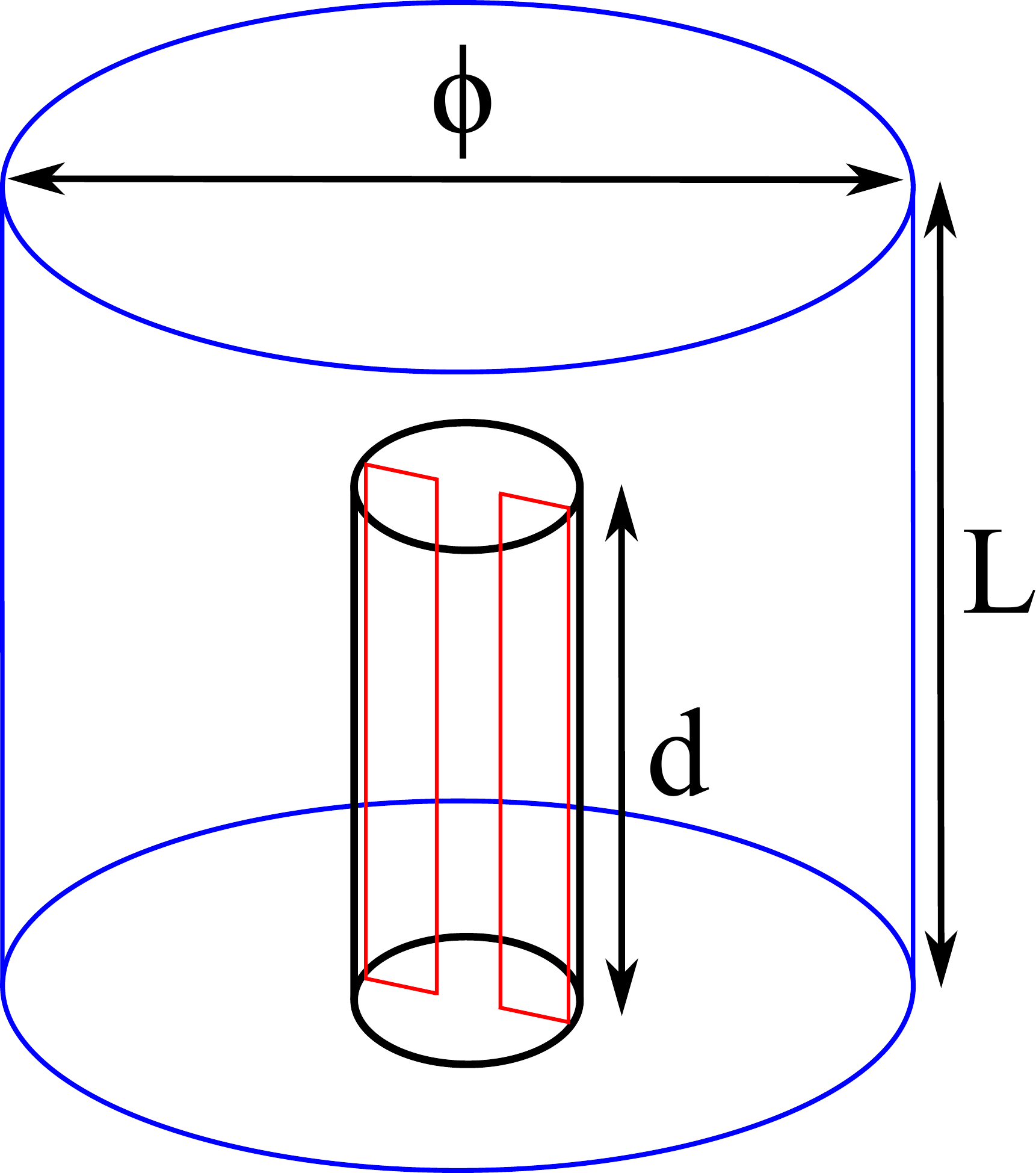}
         \caption{}
         \label{fig:CylMulticavitiesStackedInPhi_Solenoid_2cavs}
\end{subfigure}
\quad \quad \quad
\begin{subfigure}[b]{0.5\textwidth}
         \centering
         \includegraphics[width=1\textwidth]{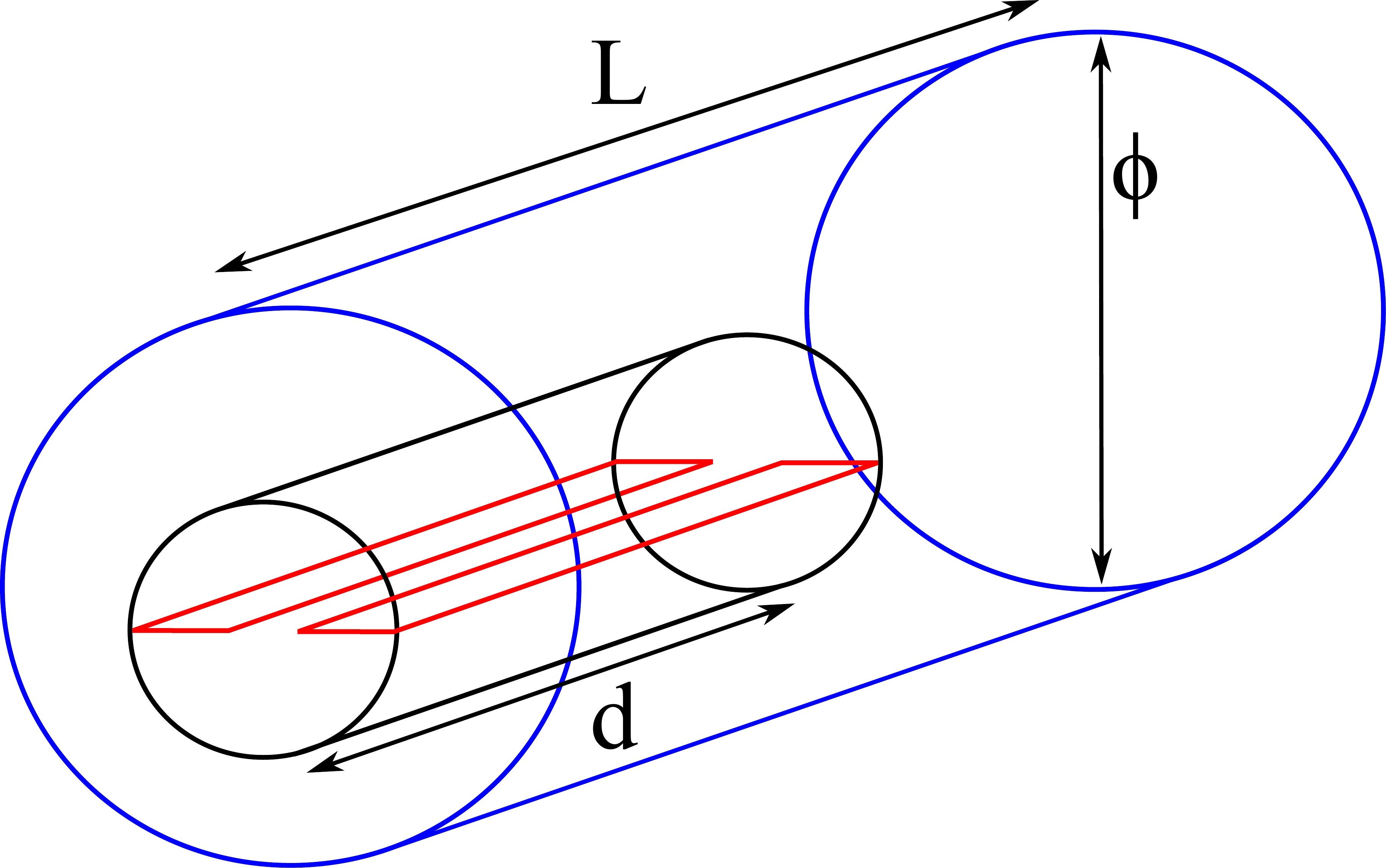}
         \caption{}
         \label{fig:CylMulticavitiesStackedInPhi_Dipole_2cavs}
\end{subfigure}
\caption{Implementation of multicavity haloscopes of two subcavities (based on a cylindrical cavity divided into two slices) stacked in $\varphi$ (a) in solenoid and (b) in dipole magnets. The red lines represent the metallic iris windows. The readout antenna for each multicavity design will be located on the upper wall of one subcavity.}
\label{fig:CylMulticavitiesStackedInPhi_Solenoid_and_Dipole_2cavs}
\end{figure}

\begin{figure}[h]
\centering
\begin{subfigure}[b]{0.22\textwidth}
         \centering
         \includegraphics[width=1\textwidth]{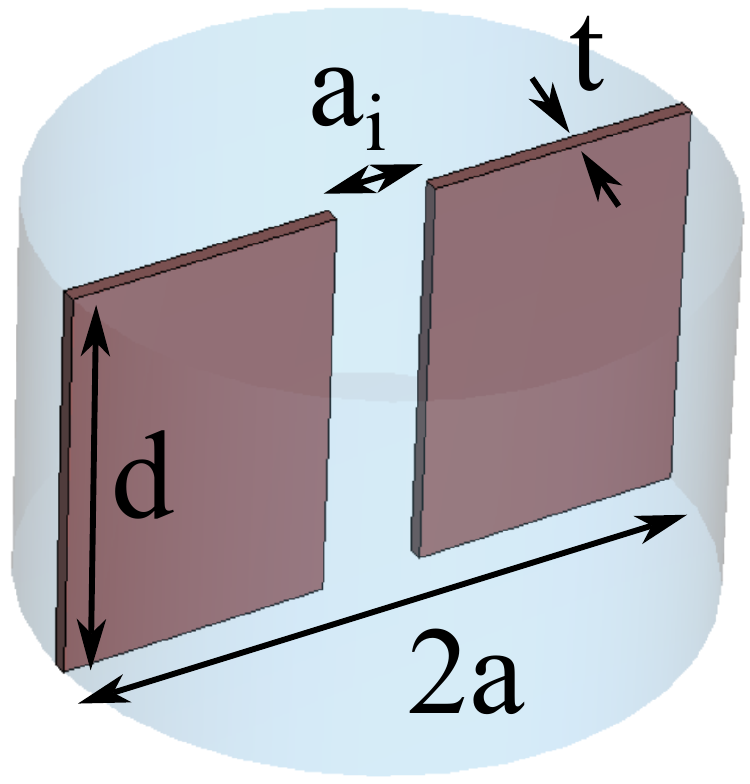}
         \caption{}
         \label{fig:CylCavsStackedInPhi_Solenoid_2subcavities_GapAtRhoPos_AtRhoCenter_Allind}
\end{subfigure}
\hfill
\begin{subfigure}[b]{0.22\textwidth}
         \centering
         \includegraphics[width=1\textwidth]{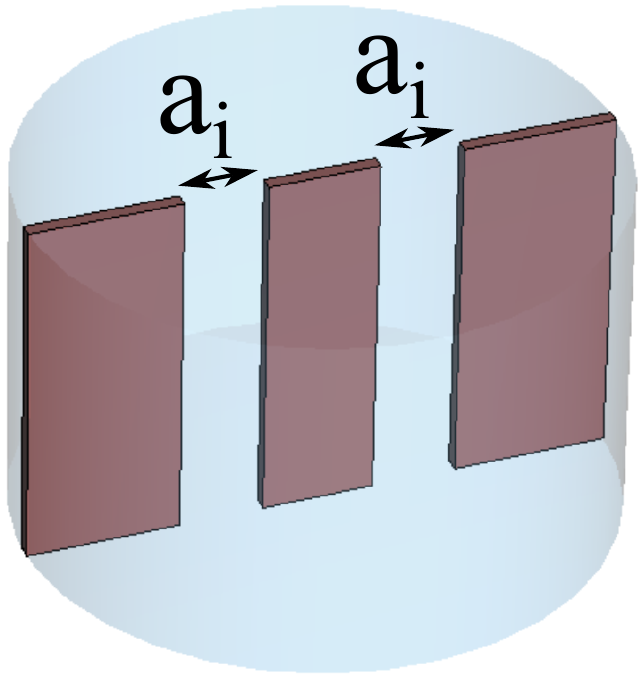}
         \caption{}
         \label{fig:CylCavsStackedInPhi_Solenoid_2subcavities_GapAtRhoPos_AtRho1-3&2-3_Allind}
\end{subfigure}
\hfill
\begin{subfigure}[b]{0.22\textwidth}
         \centering
         \includegraphics[width=1\textwidth]{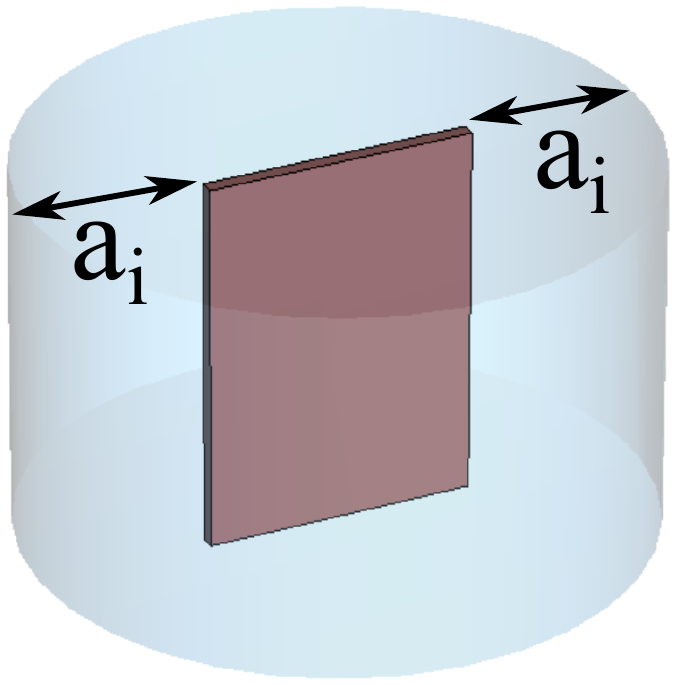}
         \caption{}
         \label{fig:CylCavsStackedInPhi_Solenoid_2subcavities_GapAtRhoPos_AtRhoEnd_Allind}
\end{subfigure}
\hfill
\begin{subfigure}[b]{0.26\textwidth}
         \centering
         \includegraphics[width=1\textwidth]{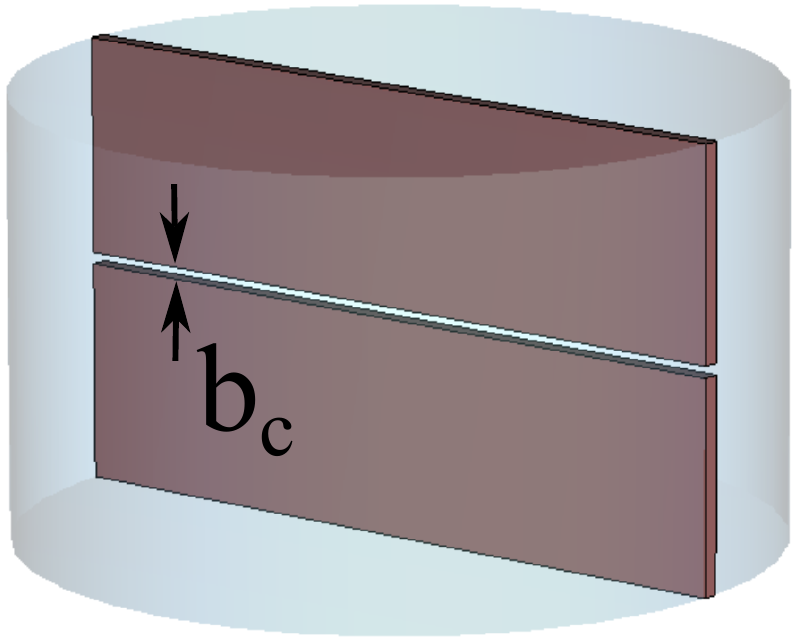}
         \caption{}
         \label{fig:CylCavsStackedInPhi_Solenoid_2subcavities_GapAtdPos_AtCenter}
\end{subfigure}
\caption{3D models of the cylindrical haloscopes based on two subcavities varying the inductive iris position along the radial axis: (a) center, (b) $1/3$ and $2/3$ of the diameter, and (c) end. (d) Case with iris in the centre, but capacitive type. $a_i$ and $b_c$ represent the inductive iris width and capacitive iris height, respectively. Light blue solids correspond with resonant cavities and brown solids with metallic iris sections. A thickness value of $t=1$~mm has been employed for all the cases.}
\label{fig:CylCavsStackedInPhi_Solenoid_2subcavities_GapAtRhoPos_and_CapCase}
\end{figure}

The initial configuration analysed in this study involves the arrangement of cylindrical subcavities along the angular axis, often referred to as the $\varphi$ axis. This arrangement's specifics are presented through visual aids, as depicted in Figures~\ref{fig:CylMulticavitiesStackedInPhi_Solenoid_2cavs} and \ref{fig:CylMulticavitiesStackedInPhi_Dipole_2cavs}. These illustrations show the spatial orientation and structural attributes of two distinct multicavity haloscope models, each incorporating two subcavities. Notably, these prototypes are designed to function within the distinct magnetic settings of solenoid ($TM_{010}$ mode) and dipole ($TE_{111}$ mode) magnets, thus exemplifying a comprehensive exploration of innovative design within diverse experimental contexts.
The CAPP group has conducted a series of investigations utilizing structures based on the first configuration, giving significant outcomes \cite{Jeong:2020cwz,Jeong:2023}. Those configurations are based on all inductive iris positioned in the centre of the cylindrical structure. Similarly, in this paper the design and analytical assessment of cylindrical haloscopes interconnecting two subcavities along the $\varphi$ axis, employing iris manipulation as a main parameter, has been carried out. This multicavity approach takes shape in various arrangements: a central iris placement (as studied by the CAPP group \cite{Jeong:2020cwz}), alongside two additional configurations with irises situated at $1/3$ and $2/3$ of the radial axis, as well as irises situated at both termini of the radial axis. In addition, as a comparison with the case of the centred iris, a design with a centred capacitive iris has been carried out. The graphical renditions of these configurations are depicted through Figure~\ref{fig:CylCavsStackedInPhi_Solenoid_2subcavities_GapAtRhoPos_and_CapCase}, showing the spatial arrangements of the subcavities and their interconnected iris elements.
\\

For the design of these structures, the following specifications have been chosen: working frequency of the axion mode $f=8.5$~GHz, and interresonator coupling $|k|=0.025$, which implies a mode clustering between the configuration modes ([+ +] and [+ \text{--}]) of $\Delta f\approx 200$~MHz (or $\sim2.4$~$\%$). The results obtained by modifying the position of the inductive iris as mentioned above are collected in Table~\ref{tab:CylCavsStackedInPhi_IndPosition}.
\begin{table}[h]
\scalebox{0.8}{
\begin{tabular}{|c|c|c|c|c|c|c|c|c|}
\hline
Type & Iris position & $2a$ (mm) & $d$ (mm) & $a_i$ or $b_c$ (mm) & $V$ (mL) & $Q_0$ & $C$ & $Q_0V^2C^2$ (L$^2$) \\ \hline\hline
Ind. & Center & $42.58$ & $26.39$ & $6.8$ & $36.77$ & $6.24\times10^4$ & $0.645$ & $35.084$ \\ \hline
Ind. & 1/3 $\&$ 2/3 & $42.55$ & $26.39$ & $6.1$ & $36.85$ & $6.19\times10^4$ & $0.665$ & $37.114$ \\ \hline
Ind. & End & $42.7$ & $26.39$ & $12$ & $37.42$ & $5.91\times10^4$ & $0.719$ & $42.772$ \\ \hline
Cap. & Center & $45$ & $26.39$ & $0.7$ & $40.88$ & $6.17\times10^4$ & $0.558$ & $32.135$ \\ \hline
Single & - & $27$ & $26.39$ & - & $15.11$ & $7.32\times10^4$ & $0.6917$ & $7.996$ \\ \hline
\end{tabular}
}
\centering
\caption{\label{tab:CylCavsStackedInPhi_IndPosition} Comparison of the operational parameters of cylindrical multicavities stacked in $\varphi$ employed for resonating at $8.5$~GHz at solenoid magnets ($TM_{010}$ mode) with $|k|=0.025$ modifying the inductive iris position in the radial axis. Each case corresponds with the cases showed in Figure~\ref{fig:CylCavsStackedInPhi_Solenoid_2subcavities_GapAtRhoPos_and_CapCase}, except for the last one, which is the single cavity designed in the previous section (see solenoid scenario in Table~\ref{tab:SingleLongQVC}).}
\end{table}
As it can be seen, although the differences are not very high, we can confirm that the results of the structure with inductive irises at the ends of the radial axis provide a better $Q_0V^2C^2$ factor compared to the other two inductive cases analysed. When comparing the inductive iris centred on the radial axis with the capacitive iris centred on the longitudinal axis, we can conclude that the former provides slight improvements.\\

The inductive and capacitive couplings chosen in this study are based on windows that fill the entire length and diameter of the cavity, respectively. However, other iris designs could be implemented to facilitate the fabrication of prototypes. One example is the replacement of these windows with one or more holes (cylindrical or rectangular in shape, for example). A detailed study of the coupling capability provided by these geometries is necessary in order to extract their practical feasibility.\\

In relation to the scenario involving the stacking of subcavities along the $\varphi$ axis, operating in $TE_{111}$ mode and intended for integration onto a dipole magnet, it has revealed a distinctive outcome. In this case, the sole feasible configuration is the example in Figure~\ref{fig:CylMulticavitiesStackedInPhi_Dipole_2cavs}, characterized by its manifestation within the XZ plane and encompassing a maximum of two subcavities. Unfortunately, the boundary conditions imposed by the inner electrical wall of the iris section have no effect on the electric field of the $TE_{111}$ mode, so the mode is the same as in the original cavity and no increase in resonant frequency is achieved. Consequently, this particular variant of haloscope design can be reasonably omitted from further consideration.

\subsection{Stacking in $\rho$}
\label{ss:1Dmulticavities_StackinginRho}

\begin{figure}[h]
\centering
\begin{subfigure}[b]{0.3\textwidth}
         \centering
         \includegraphics[width=1\textwidth]{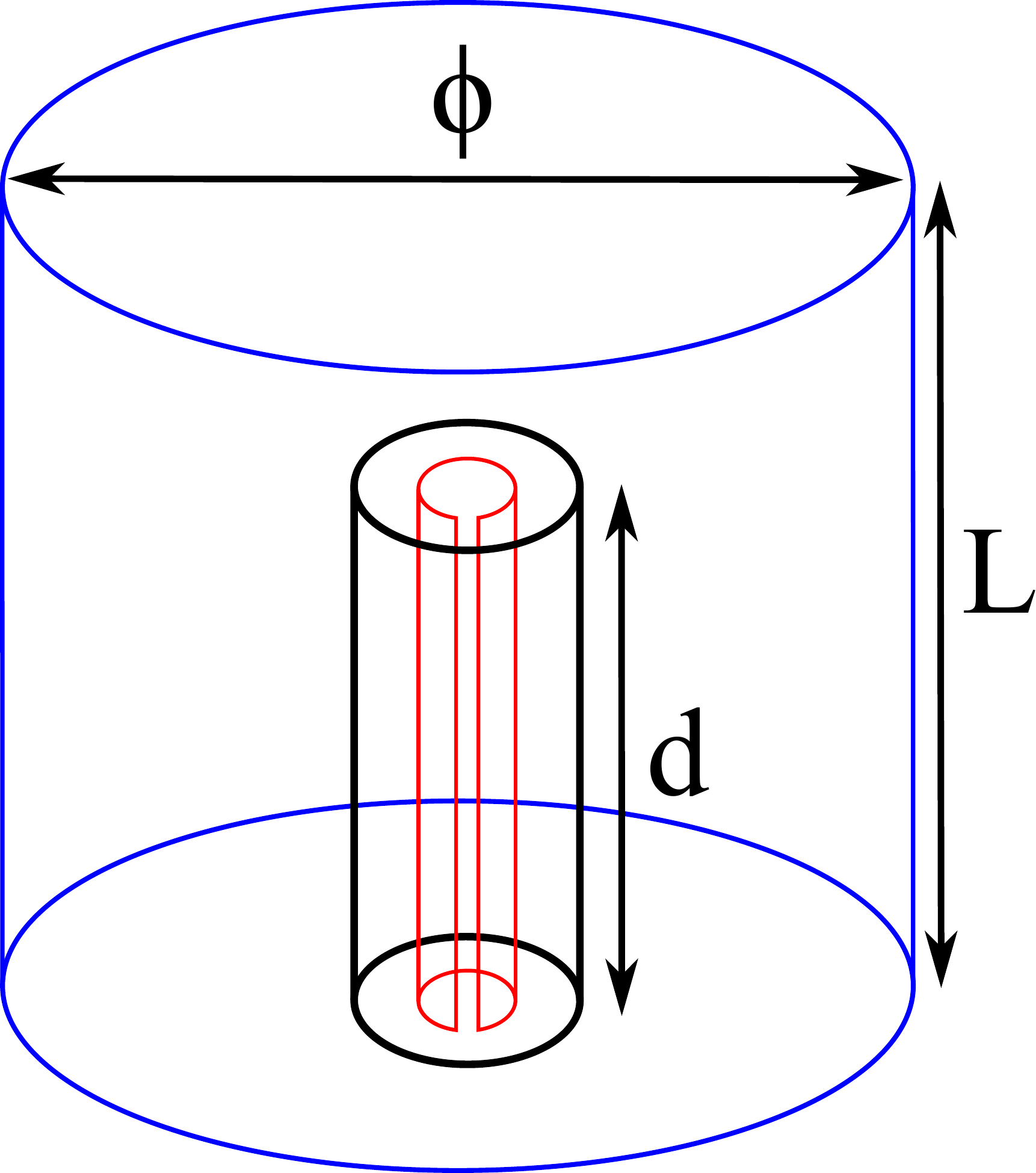}
         \caption{}
         \label{fig:CylMulticavitiesStackedInRho_Solenoid_2cavs}
\end{subfigure}
\quad \quad \quad
\begin{subfigure}[b]{0.5\textwidth}
         \centering
         \includegraphics[width=1\textwidth]{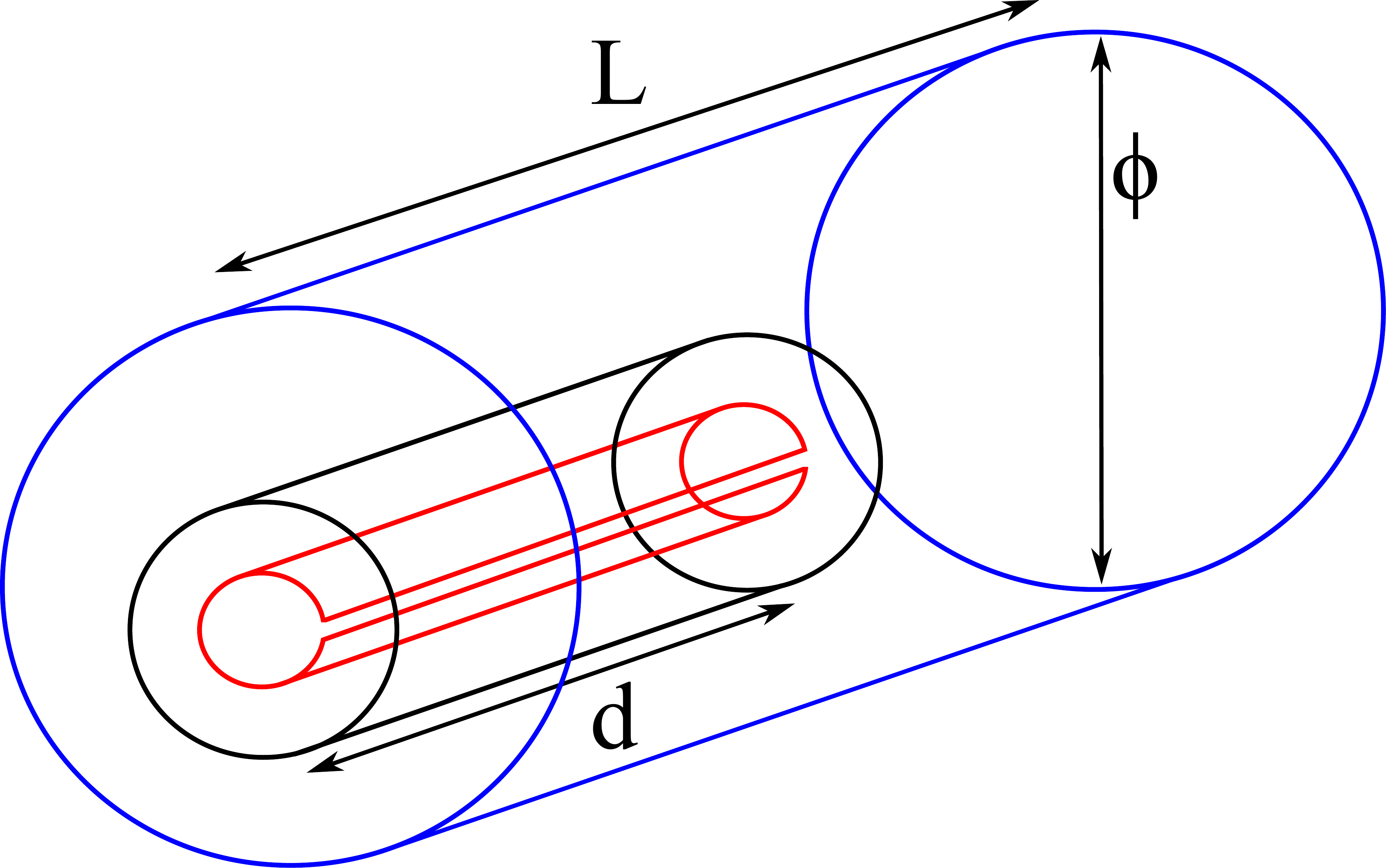}
         \caption{}
         \label{fig:CylMulticavitiesStackedInRho_Dipole_2cavs}
\end{subfigure}
\caption{Implementation of multicavity haloscopes of two subcavities stacked in $\rho$ (a) in solenoid and (b) in dipole magnets. The red lines represent the metallic iris windows.}
\label{fig:CylMulticavitiesStackedInRho_Solenoid_and_Dipole_2cavs}
\end{figure}

Illustrated in Figure~\ref{fig:CylMulticavitiesStackedInRho_Solenoid_and_Dipole_2cavs} are examples representative of both solenoids and dipoles, featuring two subcavities that are interconnected along the radial axis.
In these cases, we deal with two types of cavities: the inner one is a cylindrical cavity, an the other ones (in the case of $N$ subcavities) are coaxial cavities.\\

Notably, for the dipole-based configuration, a critical observation has emerged due to disparity in the resonance frequencies associated with the $TE_{111}$ mode between the inner and the outer subcavities. The outer (coaxial) subcavities, configured in a ring-like arrangement, necessitate substantial reduction in size along the radial axis ($\rho$), leading to a scenario where their resonance frequency cannot be harmonized with that of the inner subcavity. This limitation ultimately renders the feasibility of this particular arrangement unattainable from a practical standpoint. This option is therefore discarded again, as is the case for the $\varphi$-stacking case.\\

\begin{figure}[h]
\centering
\begin{subfigure}[b]{0.3\textwidth}
         \centering
         \includegraphics[width=1\textwidth]{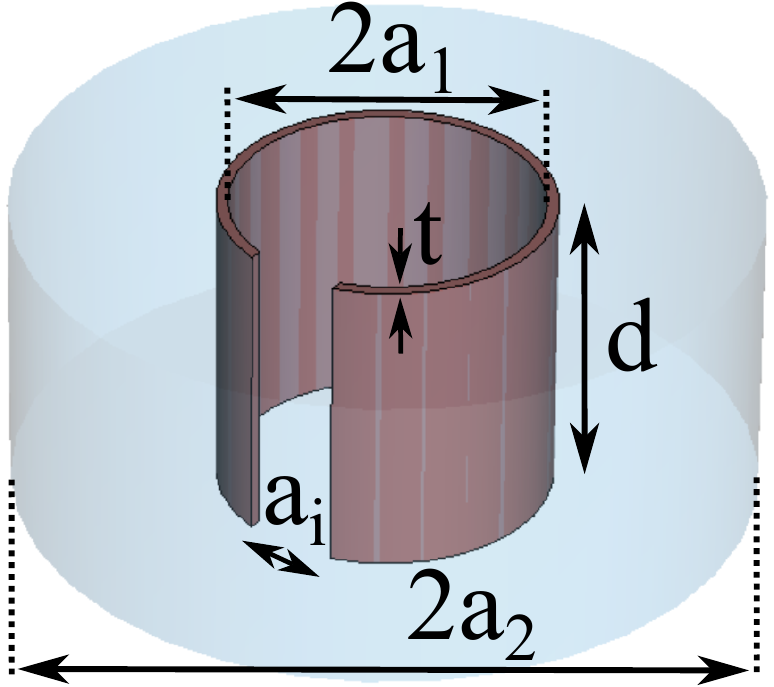}
         \caption{}
         \label{fig:CylCavsStackedInRho_Solenoid_2subcav_Allind_1iris}
\end{subfigure}
\hfill
\begin{subfigure}[b]{0.3\textwidth}
         \centering
         \includegraphics[width=1\textwidth]{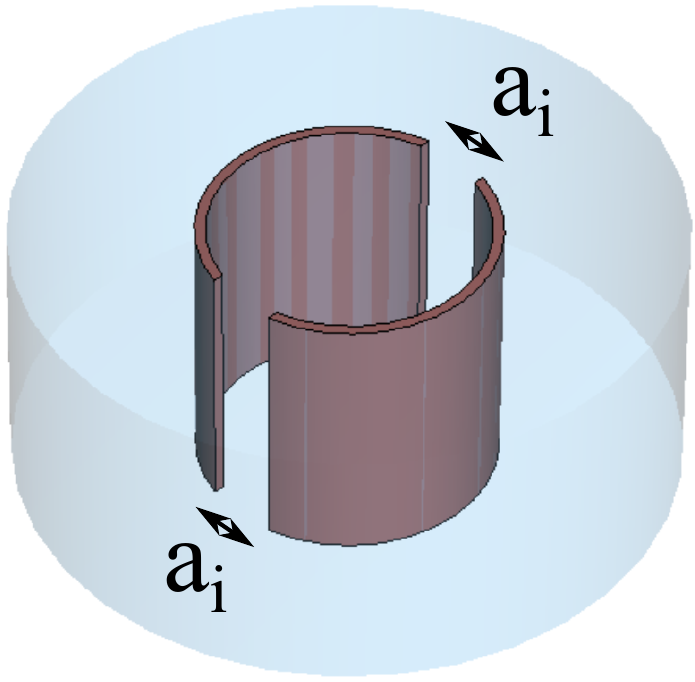}
         \caption{}
         \label{fig:CylCavsStackedInRho_Solenoid_2subcav_Allind_2iris}
\end{subfigure}
\hfill
\begin{subfigure}[b]{0.3\textwidth}
         \centering
         \includegraphics[width=1\textwidth]{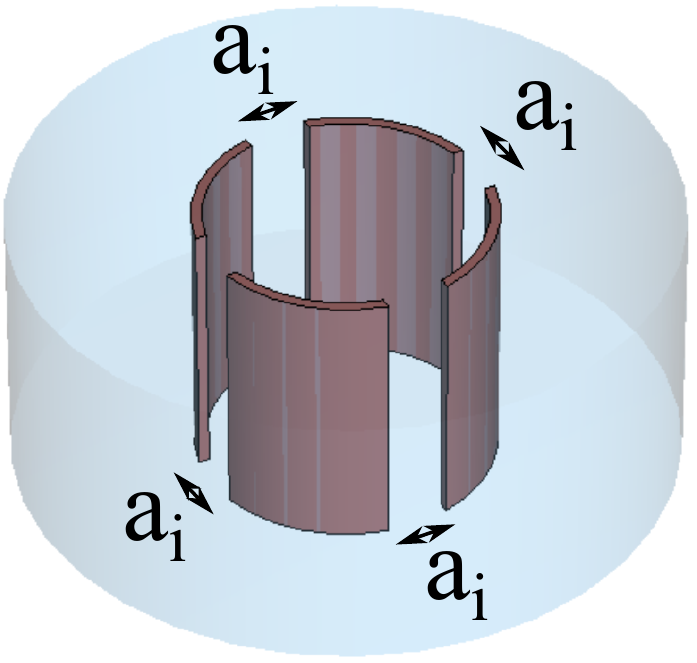}
         \caption{}
         \label{fig:CylCavsStackedInRho_Solenoid_2subcav_Allind_4iris}
\end{subfigure}
\hfill
\begin{subfigure}[b]{0.3\textwidth}
         \centering
         \includegraphics[width=1\textwidth]{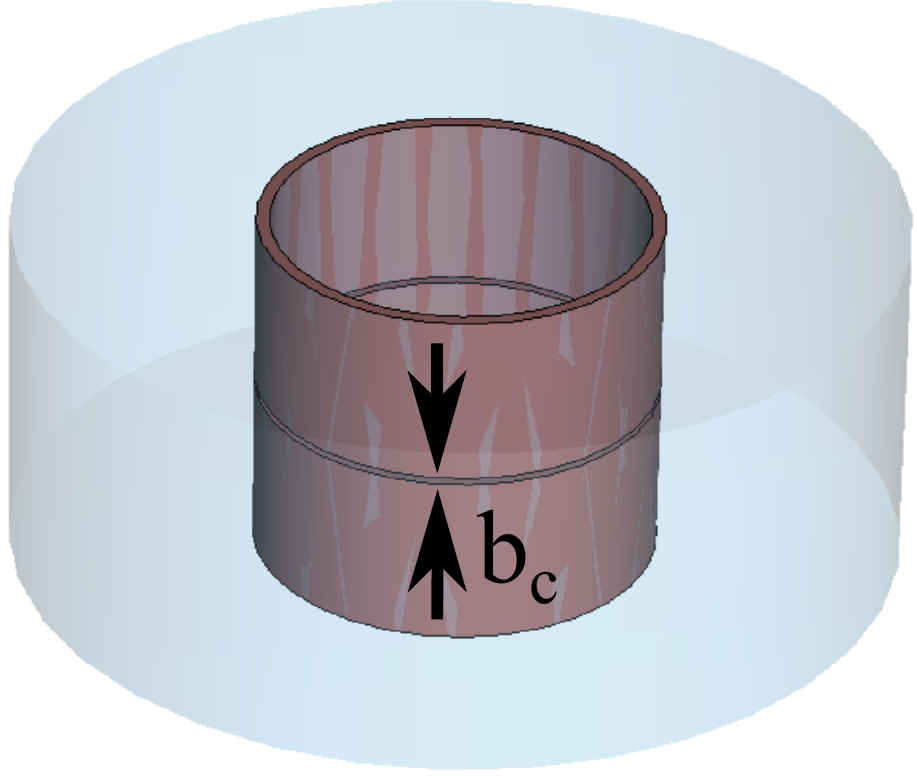}
         \caption{}
         \label{fig:CylCavsStackedInRho_Solenoid_2subcav_Allcap_1iris}
\end{subfigure}
\hfill
\begin{subfigure}[b]{0.3\textwidth}
         \centering
         \includegraphics[width=1\textwidth]{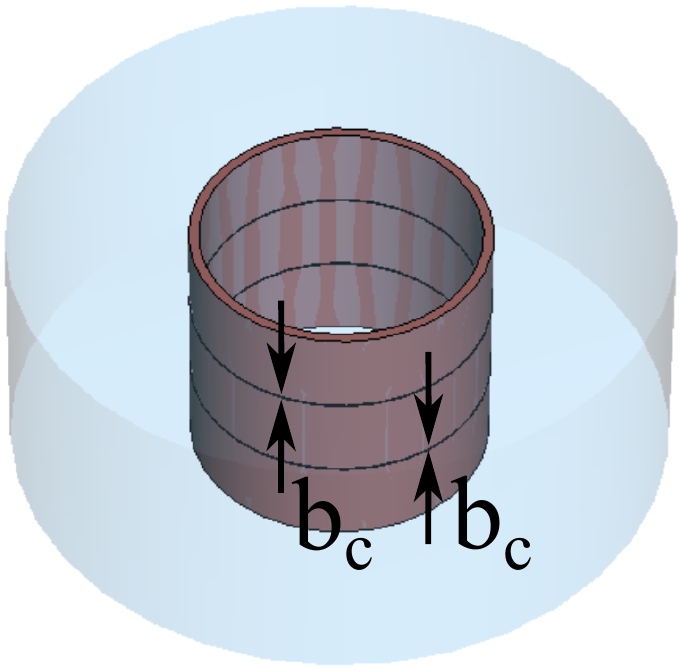}
         \caption{}
         \label{fig:CylCavsStackedInRho_Solenoid_2subcav_Allcap_2iris}
\end{subfigure}
\hfill
\begin{subfigure}[b]{0.3\textwidth}
         \centering
         \includegraphics[width=1\textwidth]{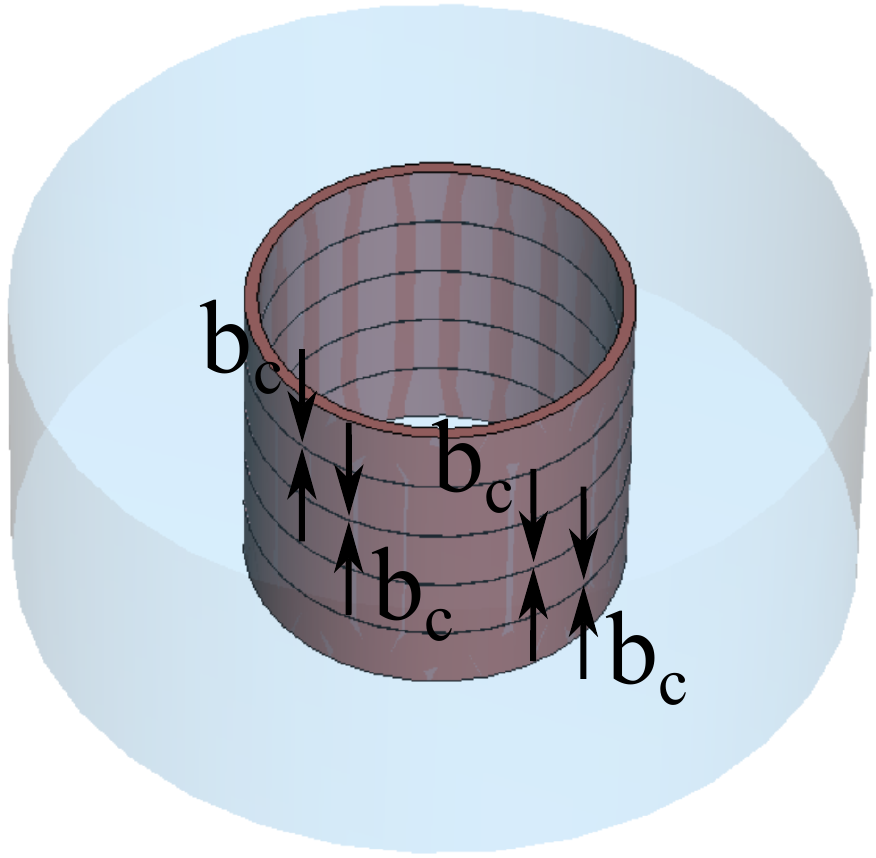}
         \caption{}
         \label{fig:CylCavsStackedInRho_Solenoid_2subcav_Allcap_4iris}
\end{subfigure}
\caption{3D models of the cylindrical haloscopes based on two subcavities varying the type (inductive and capacitive) and number of iris windows stacked in $\rho$: (a) one inductive iris, (b) two inductive iris, (c) four inductive iris, (d) one capacitive iris, (e) two capacitive iris, and (f) four capacitive iris. $a_i$ and $b_c$ represent the inductive iris width and capacitive iris height, respectively. Light blue solids correspond with resonant cavities and brown solids with metallic iris sections. A thickness value of $t=1$~mm has been employed for all the cases.}
\label{fig:CylCavsStackedInRho_Solenoid_2subcavities_IndvsCapsvsNumberIris}
\end{figure}

Regarding the solenoid configuration, specifically centered on the $TM_{010}$ mode, a thorough investigation has been conducted concerning radial ($\rho$) couplings involving two subcavities. This study encompasses the incorporation of inductive and capacitive irises, exploring three distinct scenarios contingent on the number of irises: 1, 2, and 4. Figure~\ref{fig:CylCavsStackedInRho_Solenoid_2subcavities_IndvsCapsvsNumberIris} shows the 3D models of the above-mentioned analysed structures.
\\

For the design of these structures we have set an operating frequency of $f=8.5$~GHz, and an interresonator coupling of $|k|=0.025$. Table~\ref{tab:CylCavsStackedInRho_Solenoid_2subcavities_IndvsCapsvsNumberIris} shows the outcomes of the studies modifying the type and number of the iris windows.
\begin{table}[h]
\scalebox{0.77}{
\begin{tabular}{|c|c|c|c|c|c|c|c|c|c|}
\hline
Type & $\#$ iris & $2a_1$ (mm) & $2a_2$ (mm) & $d$ (mm) & $a_i$ or $b_c$ (mm) & $V$ (mL) & $Q_0$ & $C$ & $Q_0V^2C^2$ (L$^2$) \\ \hline\hline
Ind. & 1 & $26.3$ & $62.5$ & $26.39$ & $a_i=8$ & $78.97$ & $5.75\times10^4$ & $0.582$ & $121.303$ \\ \hline
Ind. & 2 & $26.2$ & $62.3$ & $26.39$ & $a_i=7$ & $78.53$ & $5.78\times10^4$ & $0.687$ & $168.152$ \\ \hline
Ind. & 4 & $26.15$ & $62.1$ & $26.39$ & $a_i=6$ & $78.69$ & $5.96\times10^4$ & $0.65$ & $155.665$ \\ \hline
Cap. & 1 & $27.85$ & $65.7$ & $26.39$ & $b_c=0.5$ & $87.24$ & $5.78\times10^4$ & $0.666$ & $194.957$ \\ \hline
Cap. & 2 & $27.78$ & $65.55$ & $26.39$ & $b_c=0.14$ & $86.8$ & $5.9\times10^4$ & $0.663$ & $195.062$ \\ \hline
Cap. & 4 & $27.75$ & $65.5$ & $26.39$ & $b_c=0.06$ & $86.4$ & $5.92\times10^4$ & $0.649$ & $186.017$ \\ \hline
Single & - & $2a=27$ & - & $26.39$ & - & $15.11$ & $7.32\times10^4$ & $0.6917$ & $7.996$ \\ \hline
\end{tabular}
}
\centering
\caption{\label{tab:CylCavsStackedInRho_Solenoid_2subcavities_IndvsCapsvsNumberIris} Comparison of the operational parameters of cylindrical multicavities stacked in $\rho$ employed for resonating at $8.5$~GHz at solenoid magnets ($TM_{010}$ mode) with $|k|=0.025$ modifying the type and number of iris windows. Each case corresponds with the cases showed in Figure~\ref{fig:CylCavsStackedInRho_Solenoid_2subcavities_IndvsCapsvsNumberIris}, except for the last one, which is the single cavity designed in the previous section (see solenoid scenario in Table~\ref{tab:SingleLongQVC}).}
\end{table}
The obtained outcomes demonstrate increased volumetric capacities when compared to both single cavities and $\varphi$-coupled configurations, as outlined in Tables~\ref{tab:SingleLongQVC} and \ref{tab:CylCavsStackedInPhi_IndPosition}. Broadly speaking, the analyzed scenarios share analogous characteristics; nevertheless, it is pertinent to highlight that the configuration featuring a solitary inductive iris exhibits a diminished form factor due to its asymmetrical geometry. Additionally, it is affirmed that structures incorporating capacitive irises offer superior $Q_0V^2C^2$ factors. However, it is noteworthy that the latter configurations encounter certain structural complexities, the exploration of which is reserved for future investigations\footnote{The cases with 2 and 4 capacitive irises have to be slightly modified to make them manufacturable. These modifications consist of replacing the full horizontal apertures on the $\varphi$ axis by partial gaps, adding junctions between the consecutive capacitive sections in $z$ axis.}.\\

As commented in the $\varphi$-stacking case, the selection of inductive and capacitive couplings in this investigation is grounded in windows that span the entire length and $\varphi$ of the cavity, respectively. Nevertheless, as with the $\varphi$-stacked scenario, alternative iris configurations could be employed to simplify prototype fabrication. One instance involves substituting these windows with one or more perforations. An exhaustive examination of the coupling potential offered by these geometries is essential to assess their practical viability.

\subsection{Stacking in length}
\label{ss:1Dmulticavities_StackinginLength}

\begin{figure}[h]
\centering
\begin{subfigure}[b]{0.3\textwidth}
         \centering
         \includegraphics[width=1\textwidth]{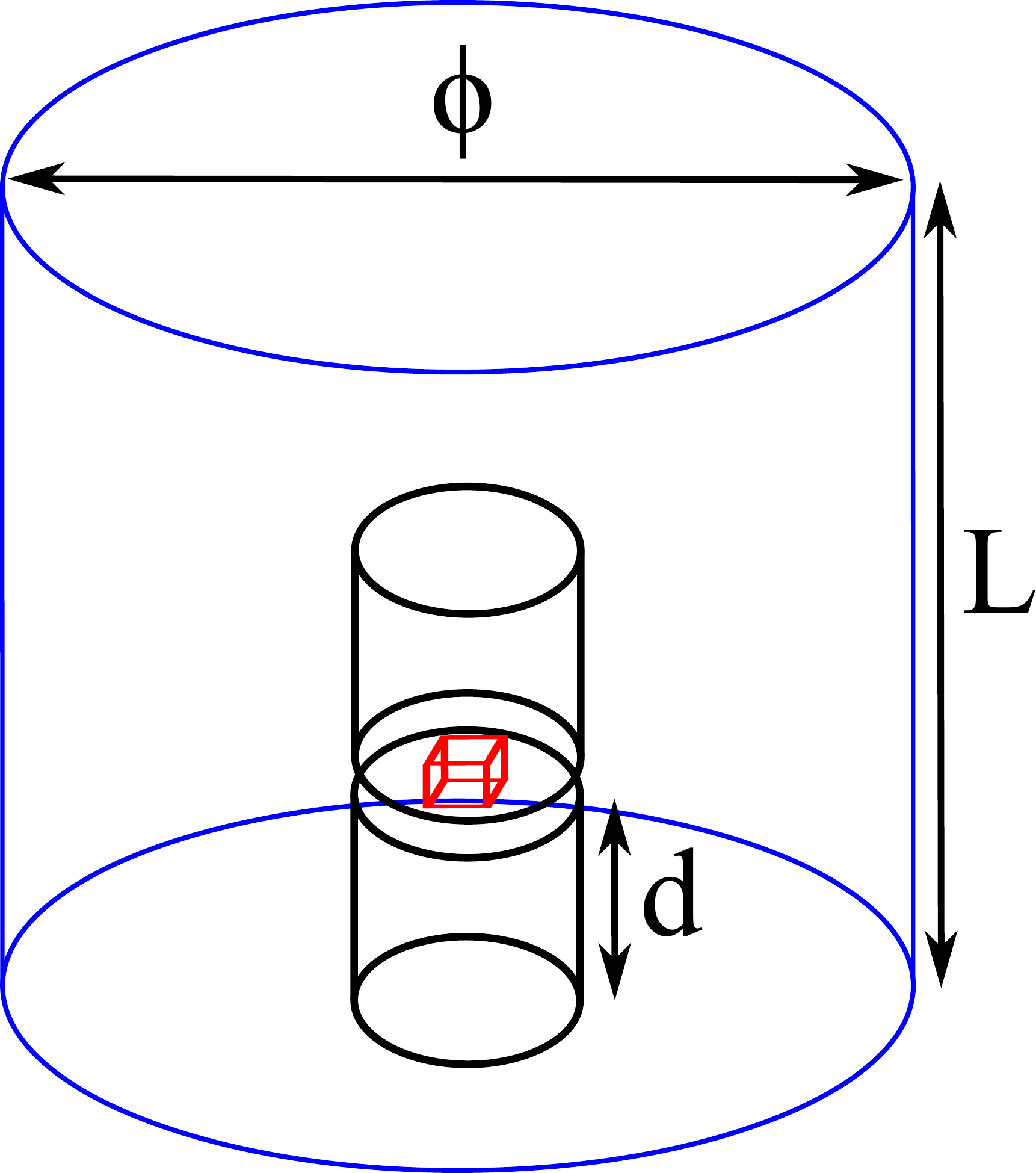}
         \caption{}
         \label{fig:CylMulticavitiesStackedInLength_Solenoid_2cavs}
\end{subfigure}
\quad \quad \quad
\begin{subfigure}[b]{0.49\textwidth}
         \centering
         \includegraphics[width=1\textwidth]{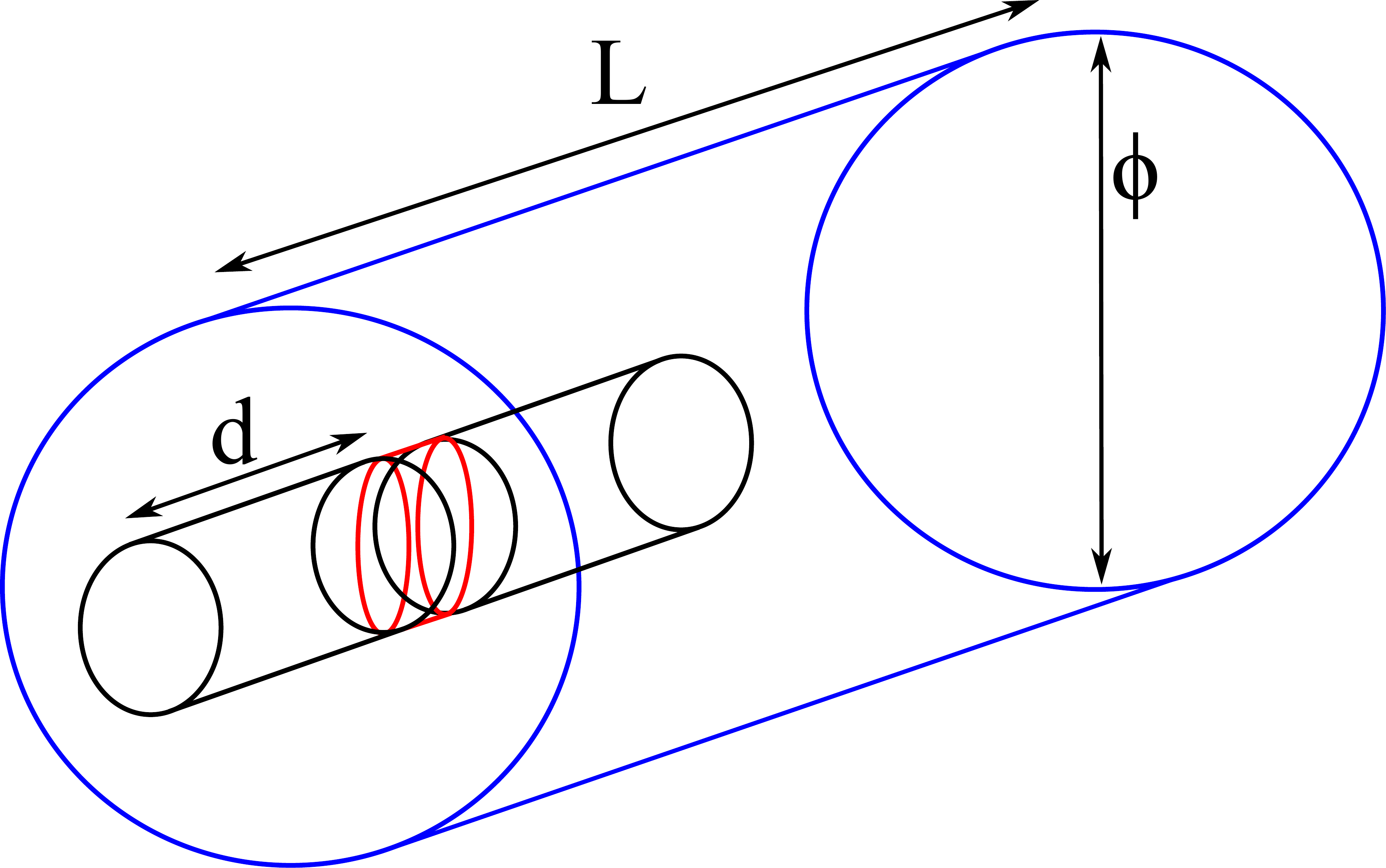}
         \caption{}
         \label{fig:CylMulticavitiesStackedInLength_Dipole_2cavs}
\end{subfigure}
\hfill
\begin{subfigure}[b]{0.18\textwidth}
         \centering
         \includegraphics[width=1\textwidth]{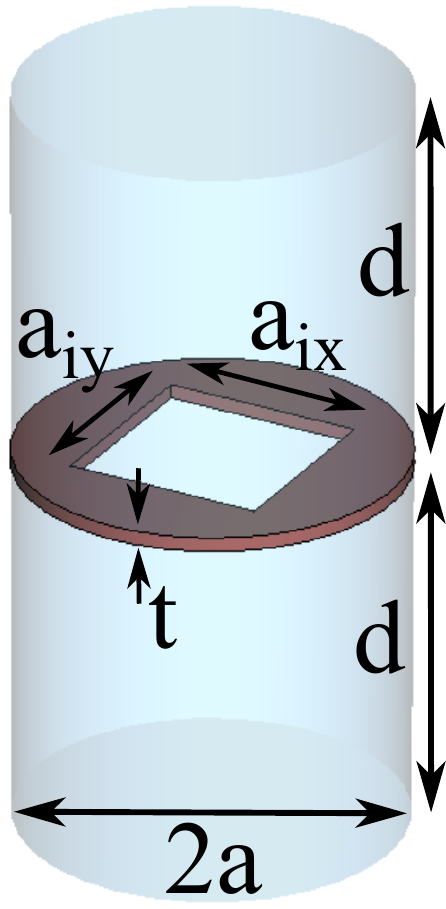}
         \caption{}
         \label{fig:CylCavsStackedInLength_Solenoid_2subcav_Allind}
\end{subfigure}
\quad \quad \quad
\begin{subfigure}[b]{0.49\textwidth}
         \centering
         \includegraphics[width=1\textwidth]{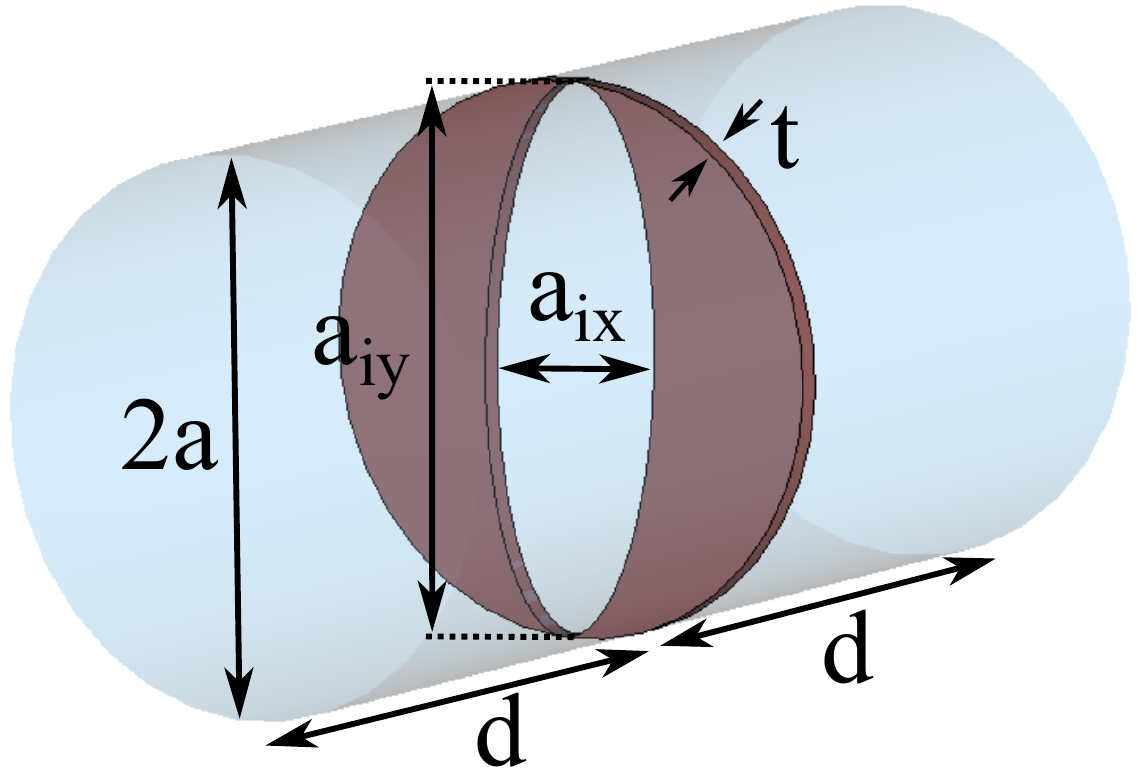}
         \caption{}
         \label{fig:CylCavsStackedInLength_Dipole_2subcav_Allind_ElliptIris}
\end{subfigure}
\caption{Implementation of two cylindrical subcavities stacked in length (a) in solenoid and (b) in dipole magnets. The red lines represent the metallic iris windows. 3D models of these cylindrical haloscopes for (c) solenoid and (d) dipole magnet. Light blue solids correspond with resonant cavities and brown solids with metallic iris sections. A thickness value of $t=1$~mm has been employed for both cases.}
\label{fig:CylMulticavitiesStackedInLength_Solenoid_and_Dipole_2cavs_and_DesignsWithAllind}
\end{figure}

Regarding the situation of length-coupled ($z$ axis) configurations (as observed in the instances with solenoid and dipole magnets in Figures~\ref{fig:CylMulticavitiesStackedInLength_Solenoid_2cavs} and \ref{fig:CylMulticavitiesStackedInLength_Dipole_2cavs}, respectively), the examination closely mirrors the approach undertaken for rectangular designs as documented in \cite{Volume_paper}.
\\

To validate the viability of cylindrical configurations, a demonstration has been presented through the design (with $f=8.5$~GHz and $|k|=0.025$) of two multicavity haloscopes, each consisting of a pair of subcavities. These designs address both solenoid and dipole magnets, integrating an inductive iris as depicted in Figures~\ref{fig:CylCavsStackedInLength_Solenoid_2subcav_Allind} and \ref{fig:CylCavsStackedInLength_Dipole_2subcav_Allind_ElliptIris}, respectively. In the case of the solenoid setup ($TM_{010}$ mode), a rectangular iris has been implemented, while the dipole configuration ($TE_{111}$ mode) incorporates an elliptical iris, which gives better results in quality factor and manufacturing possibilities \cite{Zhu:2017}. The outcomes of the simulations are illustrated in Table~\ref{tab:CylCavsStackedInz_Solenoid_and_Dipole_2subcavities_Allind}. They reveal consistently favorable results for both scenarios. The achieved $Q_0V^2C^2$ factor is over threefold superior compared to outcomes from conventional single cavity designs, demonstrating the enhancement brought about by the multicavity approach in cylindrical structures. When considering the operation in the context of a solenoid magnet and the $TM_{010}$ mode, outcomes are slightly worse than those observed in the $\varphi$ and $\rho$ stacking configurations. Nevertheless, unlike these two stacking methods, length ($z$) coupling remains a feasible option for the $TE_{111}$ mode when operating within a dipole magnet configuration.
\begin{table}[h]
\scalebox{0.77}{
\begin{tabular}{|c|c|c|c|c|c|c|c|c|}
\hline
Type & $2a$ (mm) & $d$ (mm) & $a_{ix}$ (mm) & $a_{iy}$ (mm) & $V$ (mL) & $Q_0$ & $C$ & $Q_0V^2C^2$ (L$^2$) \\ \hline\hline
Solenoid ($z$ stacking) & $27.1$ & $26.39$ & $14$ & $14$ & $30.44$ & $7.44\times10^4$ & $0.691$ & $32.953$ \\ \hline
Dipole ($z$ stacking) & $26.4$ & $26.39$ & $4.8$ & $26.4$ & $28.64$ & $7.4\times10^4$ & $0.709$ & $30.59$ \\ \hline
Solenoid (sing. cav.) & $27$ & $26.39$ & - & - & $15.11$ & $7.32\times10^4$ & $0.6917$ & $7.996$ \\ \hline
Dipole (sing. cav.) & $27.788$ & $26.39$ & - & - & $16.01$ & $7.89\times10^4$ & $0.6783$ & $9.299$ \\ \hline
\end{tabular}
}
\centering
\caption{\label{tab:CylCavsStackedInz_Solenoid_and_Dipole_2subcavities_Allind} Comparison of the operational parameters of cylindrical multicavities stacked in $z$ employed for resonating at $8.5$~GHz at solenoid ($TM_{010}$ mode) and dipole ($TE_{111}$ mode) magnets with $|k|=0.025$ with an inductive iris. Each case corresponds with the cases showed in Figures~\ref{fig:CylCavsStackedInLength_Solenoid_2subcav_Allind} and \ref{fig:CylCavsStackedInLength_Dipole_2subcav_Allind_ElliptIris}, except for the last two ones, which are the single cavities designed for solenoid and dipole scenarios in the previous section (see Table~\ref{tab:SingleLongQVC}).}
\end{table}

\subsection{Possibilities for increasing the subcavity volume}
\label{ss:1Dmulticavities_HigherVolume}

Furthermore, alongside the investigation of subcavity stacking within cylindrical waveguide configurations, an extended exploration has been conducted involving multicavity arrangements characterized by large dimensions. This expansion aims to facilitate an augmentation of volume and, in some cases, the improvement of quality factors, consequently amplifying the $Q_0V^2C^2$ factor. To ensure compatibility in terms of spatial dimensions, the analyses have been based on the largest magnets examined in the prior research \cite{Volume_paper}. Specifically, the focal points of this study encompass the quasi-dipole BabyIAXO magnet (considered here as dipole magnet, for simplicity), and the MRI (ADMX-EFR) solenoid magnet.\\

The methodology employed in these investigations has involved augmenting the dimensions of the subcavities examined in the preceding three sections (one representative case from each stacking type and magnet configuration). To achieve an increase in volume, it becomes necessary to amplify either the length ($d$) or the radius ($a$), a modification that can lead to the proximity of higher order modes. Such considerations necessitate an assessment of the mode clustering factor to prevent the diminishment of the form factor and the introduction of complexity in the computation of parameters $f_r$ and $Q_0$.

\subsubsection{Long subcavities}
\label{sss:1Dmulticavities_Long}

To exemplify the potential of volumetric expansion through length adjustments in cylindrical multicavity configurations, a series of four distinct structures were analysed. In all cases, a subcavity length of $d = 100$~mm was employed. Initially, the structure depicted in Figure~\ref{fig:CylCavsStackedInPhi_Solenoid_2subcavities_GapAtRhoPos_AtRhoCenter_Allind} was selected, a haloscope featuring two subcavities operating in the $TM_{010}$ mode situated within a solenoid magnet, and arranged in a $\varphi$-stacked configuration with a centrally positioned inductive iris. Subsequently, the design process was extended to encompass the structure illustrated in Figure~\ref{fig:CylCavsStackedInRho_Solenoid_2subcav_Allind_2iris}, characterized by two elongated subcavities operating in a solenoid magnet stacked along the $\rho$ axis and equipped with two inductive irises. Lastly, the design of the two structures presented in Figures~\ref{fig:CylCavsStackedInLength_Solenoid_2subcav_Allind} and \ref{fig:CylCavsStackedInLength_Dipole_2subcav_Allind_ElliptIris} (stacked along the length axis) was undertaken, representing both solenoid and dipole configurations, respectively.\\

The outcomes ensuing from the application of a $100$~mm subcavity length and the subsequent reoptimization of dimensions to sustain operation at $8.5$~GHz with $|k|=0.025$ for the first two scenarios (Figures~\ref{fig:CylCavsStackedInPhi_Solenoid_2subcavities_GapAtRhoPos_AtRhoCenter_Allind} and \ref{fig:CylCavsStackedInRho_Solenoid_2subcav_Allind_2iris}) are illustrated in Table~\ref{tab:CylCavsLongSubcavities_StackedPhivsRho_Solenoid}.
\begin{table}[h]
\scalebox{0.8}{
\begin{tabular}{|c|c|c|c|c|c|c|c|}
\hline
Case (stacking and magnet) & $2a$ (mm) & $d$ (mm) & $a_{i}$ (mm) & $V$ (mL) & $Q_0$ & $C$ & $Q_0V^2C^2$ (L$^2$) \\ \hline\hline
$\varphi$ and solenoid (Figure~\ref{fig:CylCavsStackedInPhi_Solenoid_2subcavities_GapAtRhoPos_AtRhoCenter_Allind}) & $42.63$ & $100$ & $7$ & $139.67$ & $7.98\times10^4$ & $0.648$ & $652.381$ \\ \hline
$\rho$ and solenoid (Figure~\ref{fig:CylCavsStackedInRho_Solenoid_2subcav_Allind_2iris}) & $2a_1=26.2$ & $100$ & $7.8$ & $300.15$ & $7.11\times10^4$ & $0.668$ & $2857.09$ \\
 & $2a_2=62.3$ & & & & & & \\ \hline
\end{tabular}
}
\centering
\caption{\label{tab:CylCavsLongSubcavities_StackedPhivsRho_Solenoid} Comparison of the operational parameters of cylindrical long multicavities for stacking in $\varphi$, $\rho$ and $z$. For the last one, both solenoid and dipole scenarios are shown. The designs are optimzed for resonating at $8.5$~GHz with $|k|=0.025$ with inductive iris windows.}
\end{table}
The outcomes achieved demonstrate that augmenting the length of these configurations by a factor of 4 has yielded a notable enhancement of the $Q_0V^2C^2$ factor, exceeding 18-fold improvement (see Tables~\ref{tab:CylCavsStackedInPhi_IndPosition} and \ref{tab:CylCavsStackedInRho_Solenoid_2subcavities_IndvsCapsvsNumberIris}). Furthermore, in these specific cases, it has been determined that the higher-order mode $TM_{011}$ has drawn closer (owing to the extension in $d$), now exhibiting a slightly closer proximity to the axion mode than the configuration mode [+ \text{--}] of $TM_{010}$. As a result, it becomes necessary to analyse this aspect when calculating the parameter $\Delta f$ (even though, for these instances, the decrease in the form factor remains negligible due to the substantial separation between adjacent modes). This result presents a compelling and promising trajectory for the formulation of extended haloscope designs taking advantage of these stacking methodologies.\\

Regrettably, while endeavoring to implement a length of $100$~mm in the length-stacked scenarios (Figures~\ref{fig:CylCavsStackedInLength_Solenoid_2subcav_Allind} and \ref{fig:CylCavsStackedInLength_Dipole_2subcav_Allind_ElliptIris}), a constraint in the iris aperture has been identified. In the case of solenoid configuration ($TM_{010}$ mode) an upper threshold of approximately $d = 33$~mm has been established, surpassing which the iris becomes unable to widen adequately to facilitate the requisite coupling. For the dipole scenario ($TE_{111}$ mode) this statement materializes at the boundary of $d = 60$~mm. This limitation aligns quite closely with the findings in the research outlined in \cite{Volume_paper} for multicavities of rectangular shape. Therefore, once this limit in length is reached, the only way to continue augmenting in volume is by increasing the number of subcavities.

%Sería interesante sacar un estudio de optimización de [N,$d_i$] para una longitud total de haloscopio $d_t$ optimizando $\Delta f$.

\subsubsection{Large diameter subcavities}
\label{sss:1Dmulticavities_LargeDiamenter}

In contrast to the longitudinal dimension ($d$), structures operating in the $TM_{010}$ resonant mode (for solenoid magnets) necessitate a precise radius ($a$) value due to the direct dependence of the resonance frequency on this parameter. As a result, expanding this dimension is not feasible for solenoid-type configurations. However, for multicavity structures of the dipole type ($TE_{111}$ mode), the radius can indeed be augmented. As observed in the preceding sections, only the length-coupled scenario exists. To exemplify this concept, the design of a structure based on two interconnected subcavities with an elliptical iris (see Figure~\ref{fig:CylCavsStackedInLength_Dipole_2subcav_Allind_ElliptIris}) featuring significantly larger radii has been undertaken. For this case, a diameter of $2a = 100$~mm has been chosen. This structure has been configured to operate at a frequency of $8.5$~GHz with an interresonator coupling of $|k|=0.025$. The results derived from this particular design are presented in Table\ref{tab:CylCavsLargeDiameterSubcavities_StackedLength_Dipole}.
\begin{table}[h]
\scalebox{0.8}{
\begin{tabular}{|c|c|c|c|c|c|c|c|}
\hline
Case (stacking and magnet) & $2a$ (mm) & $d$ (mm) & $a_{i}$ (mm) & $V$ (mL) & $Q_0$ & $C$ & $Q_0V^2C^2$ (L$^2$) \\ \hline\hline
$z$ and dipole (Figure~\ref{fig:CylCavsStackedInLength_Dipole_2subcav_Allind_ElliptIris}) & $100$ & $17.48$ & $a_{ix}=4.8$ & $272.43$ & $6.92\times10^4$ & $0.574$ & $1693.101$ \\
 & & & $a_{iy}=100$ & & & & \\ \hline
\end{tabular}
}
\centering
\caption{\label{tab:CylCavsLargeDiameterSubcavities_StackedLength_Dipole} Operational parameters of a cylindrical multicavity with large diameter subcavities stacked in $z$ with an elliptical iris windows, operating at a dipole magnet ($TE_{111}$ mode). The prototype works at $8.5$~GHz with $|k|=0.025$.}
\end{table}
It is noteworthy to highlight the substantial decrease in the cavity length required for tuning the resonance frequency in comparison to the newly set diameter. Additionally, a considerable augmentation in volume is evident when compared to the standard diameter configuration (refer to Table~\ref{tab:CylCavsStackedInz_Solenoid_and_Dipole_2subcavities_Allind}). Consequently, for this specific situation, it can be inferred that a fourfold increase in diameter leads to a remarkable enhancement in the $Q_0V^2C^2$ by over 55 times. This outcome once again presents a promising avenue to leverage a variety of dipole-type magnets.\\

Unfortunately, an issue arises when there is the potential for TM modes to transition into $TE_{111}$ mode as the cavity radius experiences a substantial increase, particularly at high frequencies such as W-Band. In these instances, the phenomenon of mode mixing can present challenges, as the frequency crossings during tuning can become quite numerous. This situation could lead to complications in accurately determining the parameters $f_r$ and $Q_0$, and there might be a risk of diminished form factor. Therefore, when undertaking the design of structures of this nature, it becomes imperative to exercise meticulous caution and consideration regarding these mode transitions in the haloscope response.\\

Regarding structures tailored for solenoid magnets with either $\varphi$ or $\rho$ coupling, a somewhat distinct approach can be adopted. For a fully utilization of the magnet's diameter (for instance, the $800$~mm diameter of the MRI magnet \cite{Volume_paper}), one can opt for an overall haloscope diameter that aligns with that of the magnet. Subsequently, the required number of radial subcavities can be implemented to attain the desired resonant frequency. This approach is tenable due to the general trend of higher resonance frequencies corresponding to smaller volumes (as indicated by equation \ref{eq:frmnl_cyl_TM}).

\subsection{Tuning}
\label{ss:1Dmulticavities_Tuning}

The tuning mechanisms elucidated in section~\ref{ss:SingleCavities_Tuning} for single cavities can be similarly adapted for the context of cylindrical structures based on 1D multicavities. The CAPP group has conducted investigations wherein dielectric rods were employed to tune cylindrical 1D multicavities in solenoids with 2, 4, 6, and 8 subcavities coupled in the $\varphi$ direction driven by an inductive central iris \cite{Jeong:2023}.\\

\begin{figure}[h]
\centering
\includegraphics[width=1\textwidth]{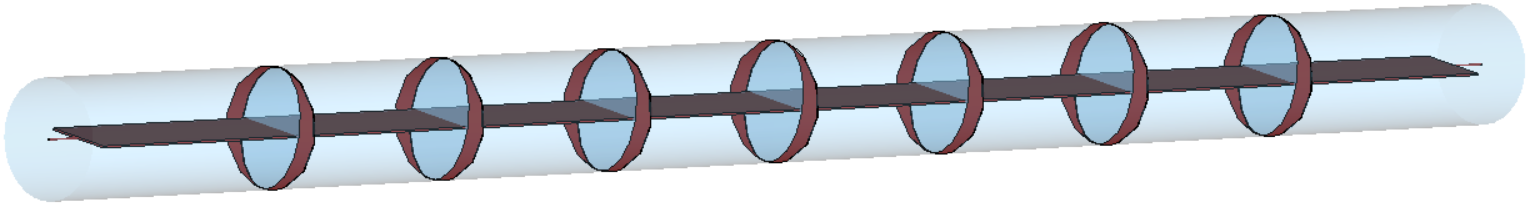}
\caption{3D model of a metallic plate system for frequency tuning implemented in a haloscope based on 8 subcavities interconnected by elliptical irises in the longitudinal axis.}
\label{fig:CylCavsStackedInLength_Dipole_8subcav_Allind_ElliptIris_PlateTuning_UHF}
\end{figure}

Furthermore, endeavors are currently in progress within the RADES group to implement the tuning systems detailed in \cite{RADES_BabyIAXO_ArXiv} (based on metal rotation plates) for 1D cylindrical multicavities operating at dipole magnets. In the same line, a haloscope operating at $213.6$~MHz (UHF-band) with the $TE_{111}$ mode (scenario for a dipole magnet) based on 8 subcavities coupled in the longitudinal axis through elliptical irises (see Figure~\ref{fig:CylCavsStackedInLength_Dipole_2subcav_Allind_ElliptIris}) has been studied where a metallic frequency tuning plate has been incorporated (see Figure~\ref{fig:CylCavsStackedInLength_Dipole_8subcav_Allind_ElliptIris_PlateTuning_UHF}).
The first simulations have provided interesting results by a total plate rotation of $60^o$ leading to a tuning range of $8$~MHz ($3.74$~$\%$), with form and quality factors within the [$0.055$, $0.321$] and [$1.31\times10^5$, $1.32\times10^5$] ranges, respectively. The development of this particular aspect of research has been assigned as a future avenue of exploration.

\section{2D and 3D multicavities}
\label{s:2D3Dmulticavities}

This study has delved into the utilization of multiple stacking axes, encompassing structures of 2D or 3D configurations. As elucidated in preceding sections, the possibility of stacking in $\varphi$ and $\rho$ arises exclusively for multicavities operating under the $TM_{010}$ mode (pertaining to the solenoid magnet context). Consequently, for 2D/3D multicavities, only this mode of operation remains tenable. As a consequence, the subsequent combinations have been subject to exploration: $\varphi$ and $\rho$ (2D), $\varphi$ and $z$ (2D), $\rho$ and $z$ (2D), and $\varphi$, $\rho$, and $z$ (3D).

\subsection{Stacking in $\varphi$ and $\rho$}
\label{ss:2D3Dmulticavities_StackinginPhiRho}

\begin{figure}[h]
\centering
\includegraphics[width=0.35\textwidth]{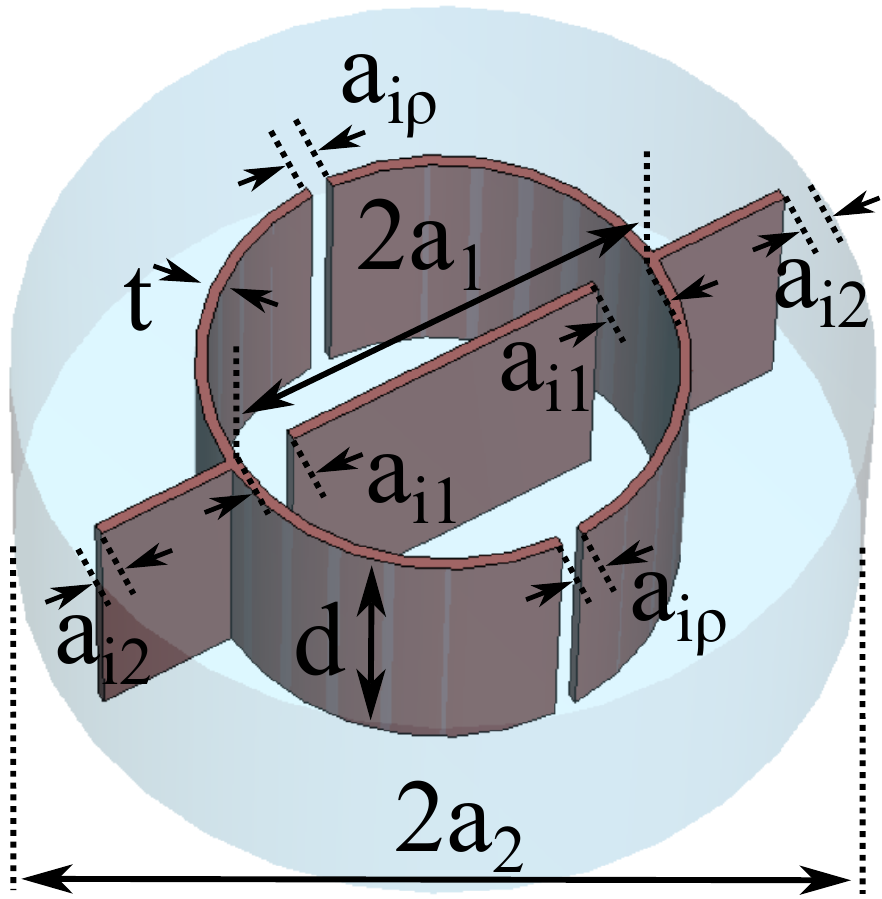}
\caption{3D model of a 2D multicavity employing $\varphi$ and $\rho$ stacking for operation at solenoid magnets. Light blue solids correspond with resonant cavities and brown solids with metallic iris sections. An iris section thickness of $t=1$~mm has been selected.}
\label{fig:CylCavsStackedInPhi2cav_and_Rho2cav_Solenoid_Allind}
\end{figure}

For this initial configuration, any permutation of inductive and capacitive irises, as expounded in sections \ref{ss:1Dmulticavities_StackinginPhi} and \ref{ss:1Dmulticavities_StackinginRho}, could be selected. To substantiate this principle, the formulation of a 2D structure has been executed as a proof of concept. To streamline the study, an all-inductive iris configuration has been reselected for the multicavity. Moreover, concerning coupling at $\varphi$ the configuration featuring two irises at the termini has been opted for (refer to Figure~\ref{fig:CylCavsStackedInPhi_Solenoid_2subcavities_GapAtRhoPos_AtRhoEnd_Allind}). Conversely, for coupling in $\rho$, a symmetric dual iris configuration has been chosen (refer to Figure~\ref{fig:CylCavsStackedInRho_Solenoid_2subcav_Allind_2iris}). The three-dimensional representation of this structure is depicted in Figure~\ref{fig:CylCavsStackedInPhi2cav_and_Rho2cav_Solenoid_Allind}.
After an optimization process modifying slightly the dimension values shown in Figure~\ref{fig:CylCavsStackedInPhi2cav_and_Rho2cav_Solenoid_Allind} the outcomes presented in Table~\ref{tab:CylCavsStackedInPhi2cav_and_Rho2cav_Solenoid_Allind} have been determined.
\begin{table}[h]
\scalebox{0.8}{
\begin{tabular}{|c|c|c|c|c|c|c|c|}
\hline
Case (stacking and magnet) & $2a$ (mm) & $d$ (mm) & $a_{i}$ (mm) & $V$ (mL) & $Q_0$ & $C$ & $Q_0V^2C^2$ (L$^2$) \\ \hline\hline
$\varphi$ $\&$ $\rho$ and solenoid & $2a_1=81.1$ & $26.39$ & $a_{i1}=9$ & $129.28$ & $5.48\times10^4$ & $0.669$ & $409.633$ \\
 & $2a_2=43.6$ & & $a_{i2}=14$ & & & & \\
 & & & $a_{i\rho}=2$ & & & & \\ \hline
\end{tabular}
}
\centering
\caption{\label{tab:CylCavsStackedInPhi2cav_and_Rho2cav_Solenoid_Allind} Operational parameters of a 2D cylindrical multicavity stacked in $\varphi$ and $\rho$ operating at a solenoid magnet ($TM_{010}$ mode).}
\end{table}
This particular configuration has been developed to operate at approximately $8.494$~GHz. A mode clustering parameter of approximately $\Delta f = 5.6$~MHz has been achieved.\\

It can be affirmed that the outcomes achieved in relation to the volume and $Q_0V^2C^2$ factor surpass, those obtained for 1D multicavities, as evidenced by the data in Tables~\ref{tab:CylCavsStackedInPhi_IndPosition} and \ref{tab:CylCavsStackedInRho_Solenoid_2subcavities_IndvsCapsvsNumberIris}.

\subsection{Stacking in $\varphi$ and length}
\label{ss:2D3Dmulticavities_StackinginPhiLength}

\begin{figure}[h]
\centering
\includegraphics[width=0.3\textwidth]{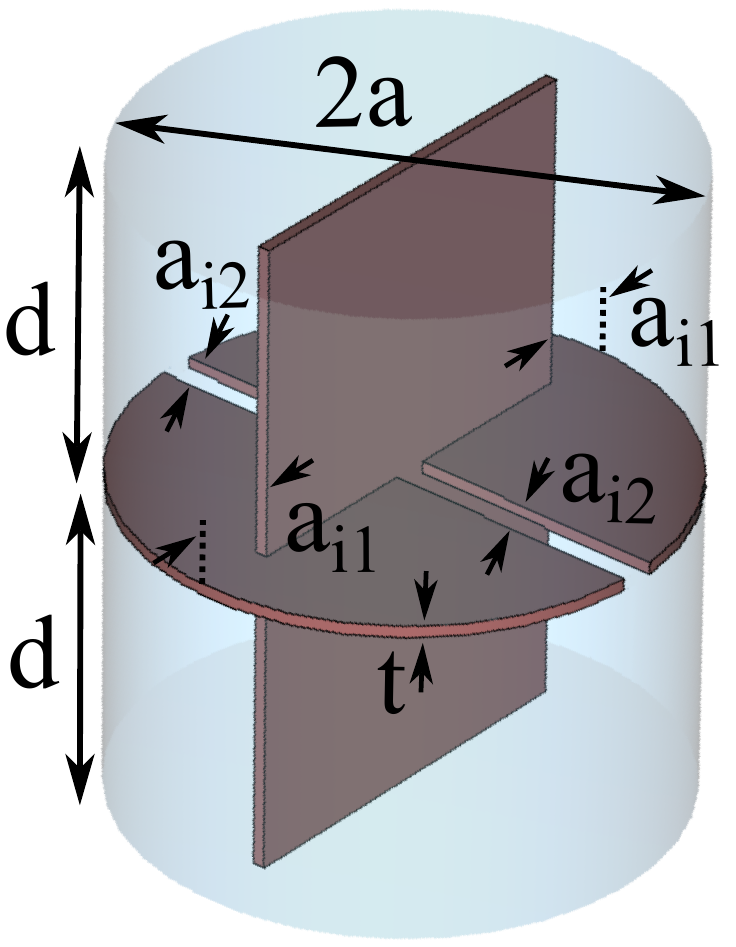}
\caption{3D model of a 2D multicavity employing $\varphi$ and $z$ stacking for operation at solenoid magnets. Light blue solids correspond with resonant cavities and brown solids with metallic iris sections. An iris section thickness of $t=1$~mm has been selected.}
\label{fig:CylCavsStackedInLength2cav_and_Phi2cav_Solenoid_Allind}
\end{figure}

An analogous approach to the previously discussed section has been undertaken for 2D cylindrical multicavities coupled along the $\varphi$ and $z$ axes. In this instance, the configuration depicted in Figure~\ref{fig:CylCavsStackedInLength2cav_and_Phi2cav_Solenoid_Allind} has been devised as the basis for the haloscope.
The design employs an all-inductive setup. When it comes to coupling along the $\varphi$ axis, dual irises have been positioned at the extremities, while for the lengthwise coupling rectangular inductive irises have been selected (similar to the scenarios depicted in Figure~\ref{fig:CylMulticavitiesStackedInLength_Solenoid_and_Dipole_2cavs_and_DesignsWithAllind}). The optimization process has yielded to the simulation results presented in Table~\ref{tab:CylCavsStackedInLength2cav_and_Phi2cav_Solenoid_Allind}.
\begin{table}[h]
\scalebox{0.8}{
\begin{tabular}{|c|c|c|c|c|c|c|c|}
\hline
Case (stacking and magnet) & $2a$ (mm) & $d$ (mm) & $a_{i}$ (mm) & $V$ (mL) & $Q_0$ & $C$ & $Q_0V^2C^2$ (L$^2$) \\ \hline\hline
$\varphi$ $\&$ $z$ and solenoid & $43.86$ & $26.39$ & $a_{i1}=6.5$ & $78.85$ & $6.58\times10^4$ & $0.646$ & $171.02$ \\
 & & & $a_{i2}=4$ & & & & \\ \hline
\end{tabular}
}
\centering
\caption{\label{tab:CylCavsStackedInLength2cav_and_Phi2cav_Solenoid_Allind} Operational parameters of a 2D cylindrical multicavity stacked in $\varphi$ and $z$ operating at a solenoid magnet ($TM_{010}$ mode).}
\end{table}
This specific arrangement has been devised to operate at approximately $8.496$~GHz. An achieved mode clustering parameter of roughly $\Delta f = 7$~MHz has been observed. Notably, the optimization stage has produced a highly favourable form factor value.\\

The results clearly indicate that the achieved $Q_0V^2C^2$ factor improvements exceed those obtained for 1D multicavities, as illustrated by the data presented in Tables~\ref{tab:CylCavsStackedInPhi_IndPosition} and \ref{tab:CylCavsStackedInz_Solenoid_and_Dipole_2subcavities_Allind}, resulting in an increase of its value by over fourfold.

\subsection{Stacking in $\rho$ and length}
\label{ss:2D3D_StackinginLengthRho}

\begin{figure}[h]
\centering
\includegraphics[width=0.4\textwidth]{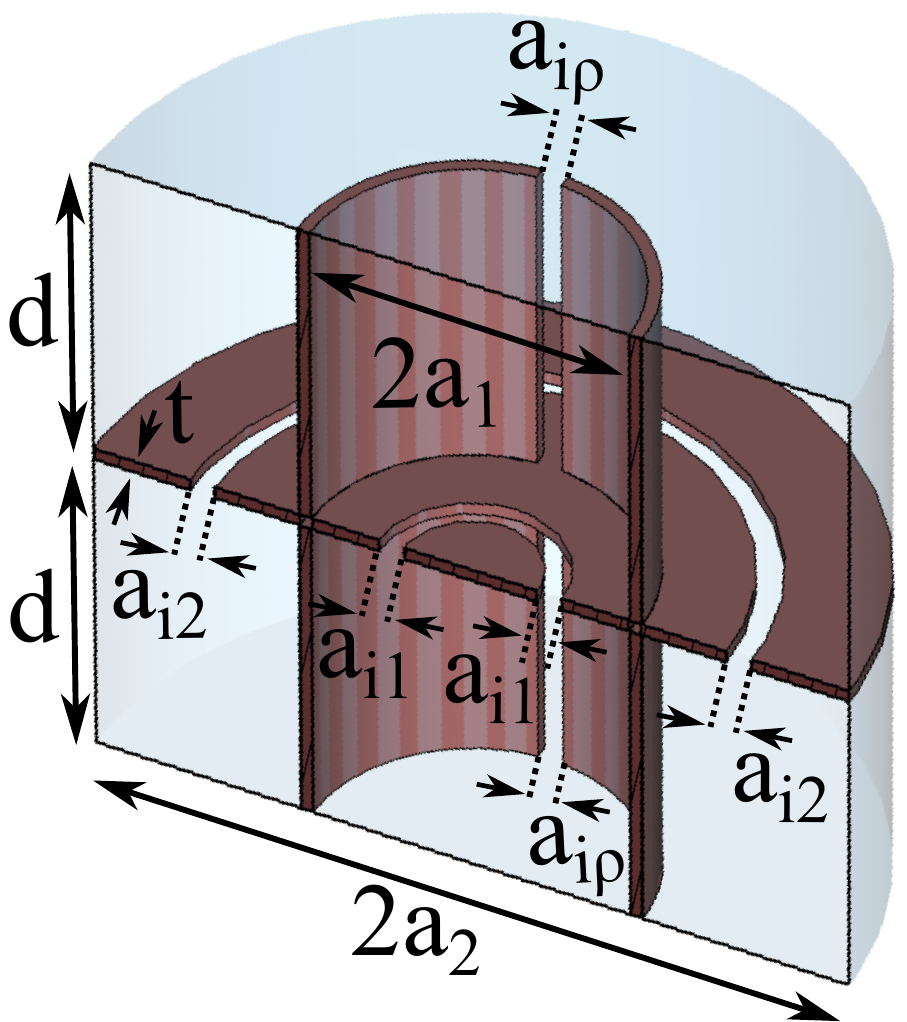}
\caption{3D model of the symmetrical half of a 2D multicavity employing $\rho$ and $z$ stacking for operation at solenoid magnets. Light blue solids correspond with resonant cavities and brown solids with metallic iris sections. An iris section thickness of $t=1$~mm has been selected.}
\label{fig:CylCavsStackedInLength2cav_and_Rho2cav_Solenoid_Allind_SymmetricalHalf}
\end{figure}

The final feasible arrangement for 2D multicavities involves the utilization of the $\rho$ and $z$ axes to stack subcavities. In this scenario, the prototype design features symmetrical inductive irises for coupling along the $\rho$ axis and ring-shaped inductive irises for coupling along the length axis, as depicted in Figure~\ref{fig:CylCavsStackedInLength2cav_and_Rho2cav_Solenoid_Allind_SymmetricalHalf}\footnote{Comparable to certain designs depicted in Figure~\ref{fig:CylCavsStackedInRho_Solenoid_2subcavities_IndvsCapsvsNumberIris}, slight adjustments to the inner inductive coupling ring of this 2D multicavity are envisioned to render it feasible, replacing the iris window by partial apertures.}.
After an optimization process modifying slightly the dimension values of the structure, the results outlined in Table~\ref{tab:CylCavsStackedInLength2cav_and_Rho2cav_Solenoid_Allind} have been obtained.
\begin{table}[h]
\scalebox{0.8}{
\begin{tabular}{|c|c|c|c|c|c|c|c|}
\hline
Case (stacking and magnet) & $2a$ (mm) & $d$ (mm) & $a_{i}$ (mm) & $V$ (mL) & $Q_0$ & $C$ & $Q_0V^2C^2$ (L$^2$) \\ \hline\hline
$\rho$ $\&$ $z$ and solenoid & $2a_1=26.95$ & $26.39$ & $a_{i1}=2$ & $165.41$ & $6.33\times10^4$ & $0.629$ & $685.84$ \\
 & $2a_2=63.95$ & & $a_{i2}=2.5$ & & & & \\
 & & & $a_{i\rho}=2.5$ & & & & \\ \hline
\end{tabular}
}
\centering
\caption{\label{tab:CylCavsStackedInLength2cav_and_Rho2cav_Solenoid_Allind} Operational parameters of a 2D cylindrical multicavity stacked in $\rho$ and $z$ operating at a solenoid magnet ($TM_{010}$ mode).}
\end{table}
This specific arrangement has been developed for operation around $8.496$~GHz. An achieved mode clustering parameter of approximately $\Delta f = 7.4$~MHz has been observed.\\

It is worth noting that the results achieved in terms of $Q_0V^2C^2$ factor outperform significantly those obtained for 1D multicavities, as indicated by the data presented in Tables~\ref{tab:CylCavsStackedInRho_Solenoid_2subcavities_IndvsCapsvsNumberIris} and \ref{tab:CylCavsStackedInz_Solenoid_and_Dipole_2subcavities_Allind}.

\subsection{Stacking in $\varphi$, $\rho$, and length}
\label{ss:2D3Dmulticavities_StackinginPhiRhoLength}

\begin{figure}[h]
\centering
\includegraphics[width=0.55\textwidth]{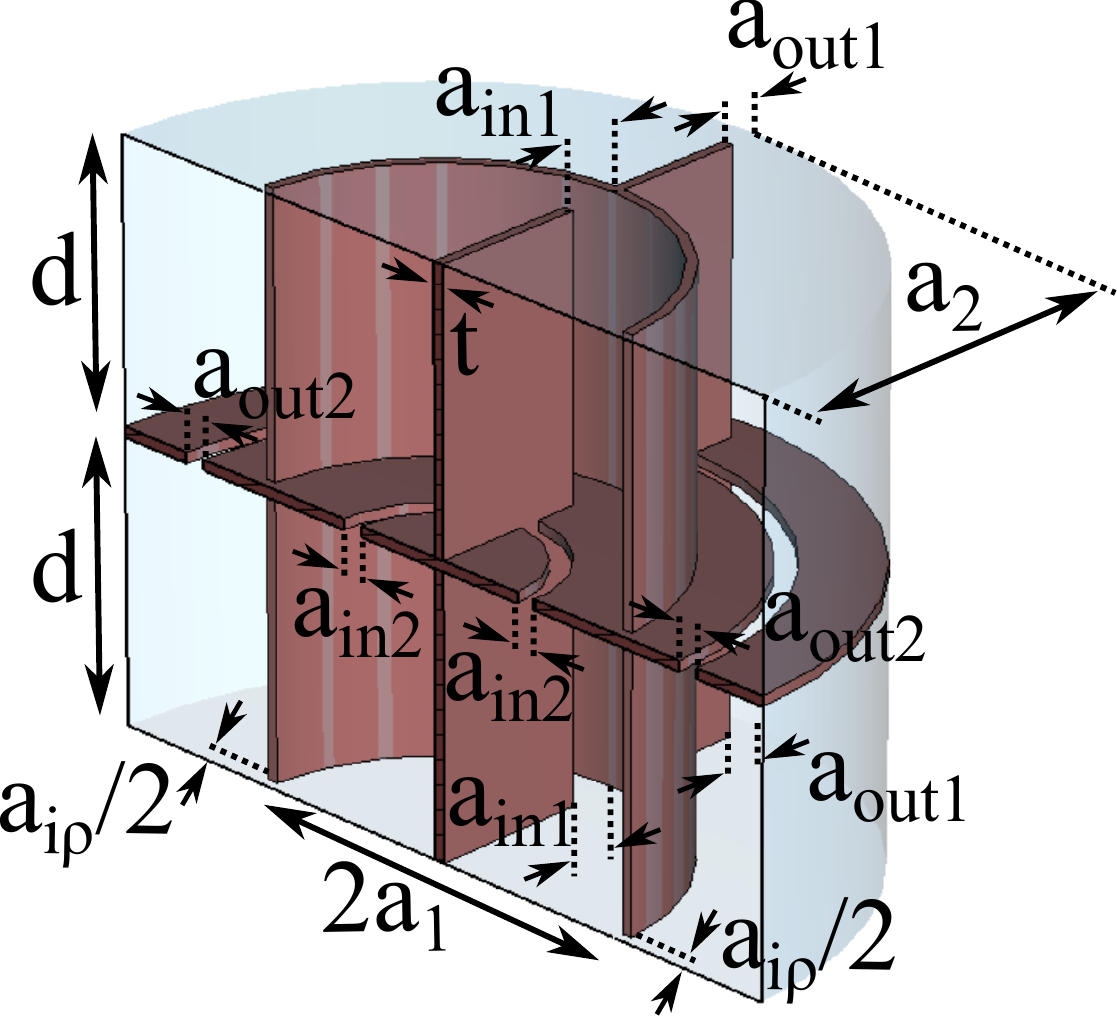}
\caption{3D model of the symmetrical half of a 3D multicavity employing $\varphi$, $\rho$, and $z$ stacking for operation at solenoid magnets. Light blue solids correspond with resonant cavities and brown solids with metallic iris sections. An iris section thickness of $t=1$~mm has been selected.}
\label{fig:CylCavsStackedInLength2cav_and_Phi2cav_and_Rho2cav_Solenoid_Allind_SymmetricalHalf}
\end{figure}

In order to construct 3D multicavity structures, the integration of various stacking orientations among subcavities is essential, encompassing coupling along the $\varphi$, $\rho$, and $z$ axes. A prototype design featuring eight subcavities incorporating all inductive irises has been formulated to serve as a demonstration of the viability of such haloscopes. In the context of $\varphi$ coupling, the selection involves two irises situated at the extremities. For the $\rho$ stacking, a pair of symmetrical irises is employed. Lastly, coupling along the longitudinal axis involves the utilization of two ring-shaped iris windows, as depicted in Figure~\ref{fig:CylCavsStackedInLength2cav_and_Phi2cav_and_Rho2cav_Solenoid_Allind_SymmetricalHalf}.
As evident, the quantity of parameters requiring optimization has substantially expanded, consequently making the design process for this structure potentially complex. However, after a fine optimization, the simulation results detailed in Table~\ref{tab:CylCavsStackedInLength2cav_and_Phi2cav_and_Rho2cav_Solenoid_Allind} have been achieved.
\begin{table}[h]
\scalebox{0.8}{
\begin{tabular}{|c|c|c|c|c|c|c|c|}
\hline
Case (stacking and magnet) & $2a$ (mm) & $d$ (mm) & $a_{i}$ (mm) & $V$ (mL) & $Q_0$ & $C$ & $Q_0V^2C^2$ (L$^2$) \\ \hline\hline
$\varphi$, $\rho$ $\&$ $z$ and solenoid & $2a_1=43.6$ & $26.39$ & $a_{in1}=9$ & $264.51$ & $6.08\times10^4$ & $0.553$ & $1299.02$ \\
 & $2a_2=81.1$ & & $a_{out1}=14$ & & & & \\
 & & & $a_{i\rho}=2$ & & & & \\
 & & & $a_{in2}=2$ & & & & \\
 & & & $a_{out2}=2.5$ & & & & \\ \hline
\end{tabular}
}
\centering
\caption{\label{tab:CylCavsStackedInLength2cav_and_Phi2cav_and_Rho2cav_Solenoid_Allind} Operational parameters of a 3D cylindrical multicavity stacked in $\varphi$, $\rho$, and $z$ operating at a solenoid magnet ($TM_{010}$ mode).}
\end{table}
This particular configuration has been designed for operation at approximately $8.501$~GHz. A mode clustering parameter of around $\Delta f = 4$~MHz has been attained.\\

Upon comparing these findings with the examined 2D structures (see Tables~\ref{tab:CylCavsStackedInPhi2cav_and_Rho2cav_Solenoid_Allind}, \ref{tab:CylCavsStackedInLength2cav_and_Phi2cav_Solenoid_Allind} and \ref{tab:CylCavsStackedInLength2cav_and_Rho2cav_Solenoid_Allind}), it is evident that the acquired $Q_0V^2C^2$ factor is better, confirming the feasibility of 3D geometries in cylindrical multicavity haloscopes.

\subsection{Hexagonal multicavities}
\label{ss:2D3Dmulticavities_HexagonalMulticavities}

Another concept briefly introduced consists of the employment of resonant cavities with hexagonal section for the constitution of 2D and 3D multicavity haloscope structures. This hexagonal spatial arrangement is distinguished as an optimally efficient approach for populating structures with circular geometry, thus optimizing the volumetric occupancy. The articulation of this notion has been previously introduced in a prior publication \cite{Bowring:2018} wherein hexagonal single cavities positioned adjacently within a common magnet, to be in-phase added by means of power combiners, were subjected to analysis and evaluation.\\

\begin{figure}[h]
\centering
\begin{subfigure}[b]{0.2\textwidth}
         \centering
         \includegraphics[width=1\textwidth]{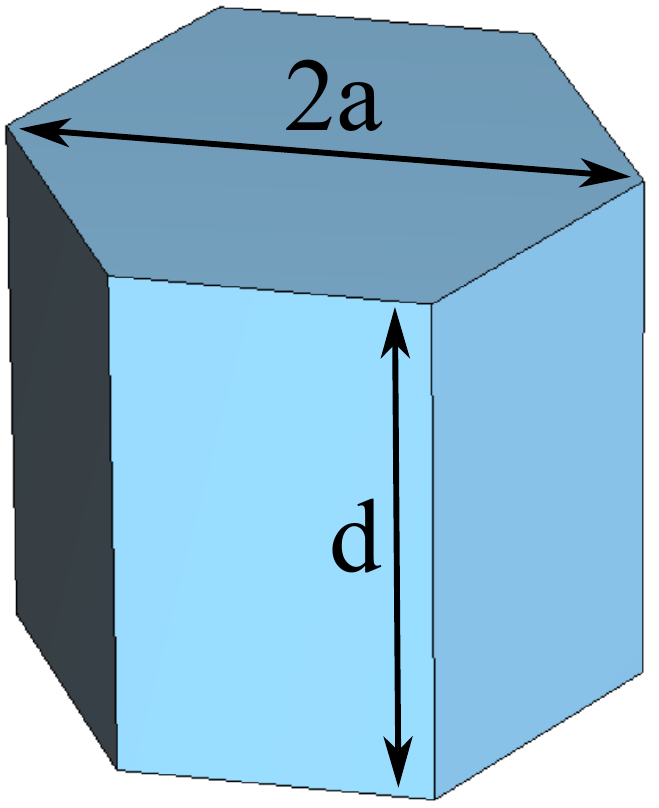}
         \caption{}
         \label{fig:HexCav}
\end{subfigure}
\hfill
\begin{subfigure}[b]{0.32\textwidth}
         \centering
         \includegraphics[width=1\textwidth]{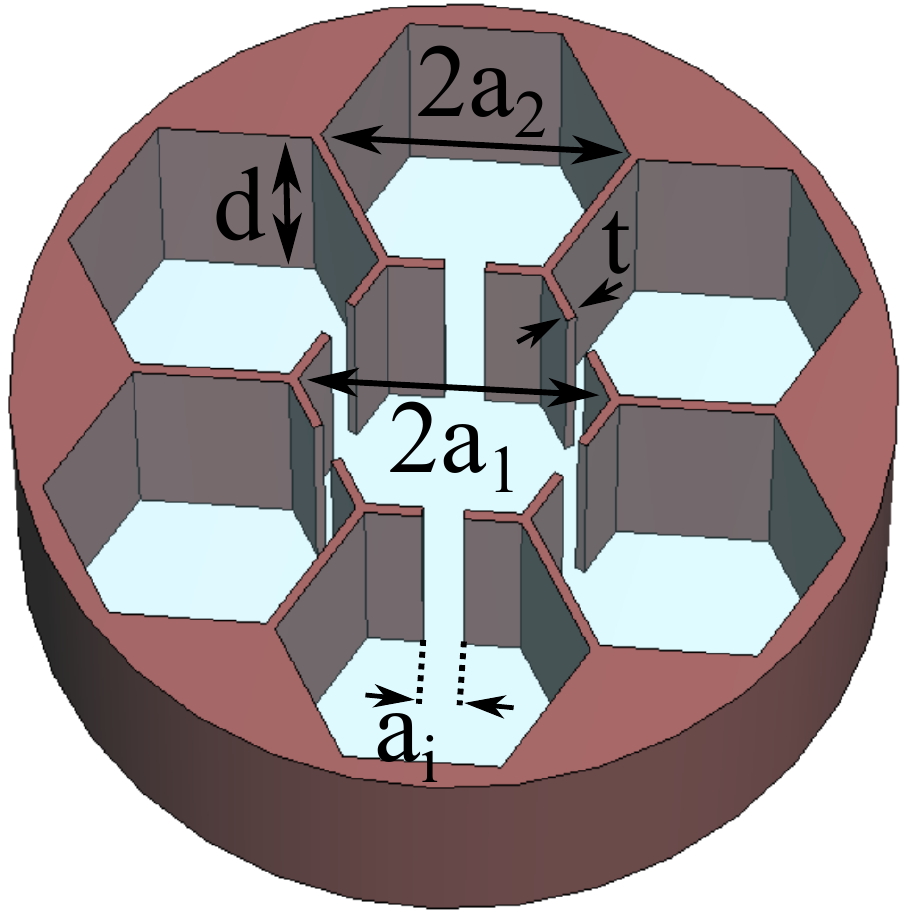}
         \caption{}
         \label{fig:HexCav_2D_7subcavities}
\end{subfigure}
\hfill
\begin{subfigure}[b]{0.4\textwidth}
         \centering
         \includegraphics[width=1\textwidth]{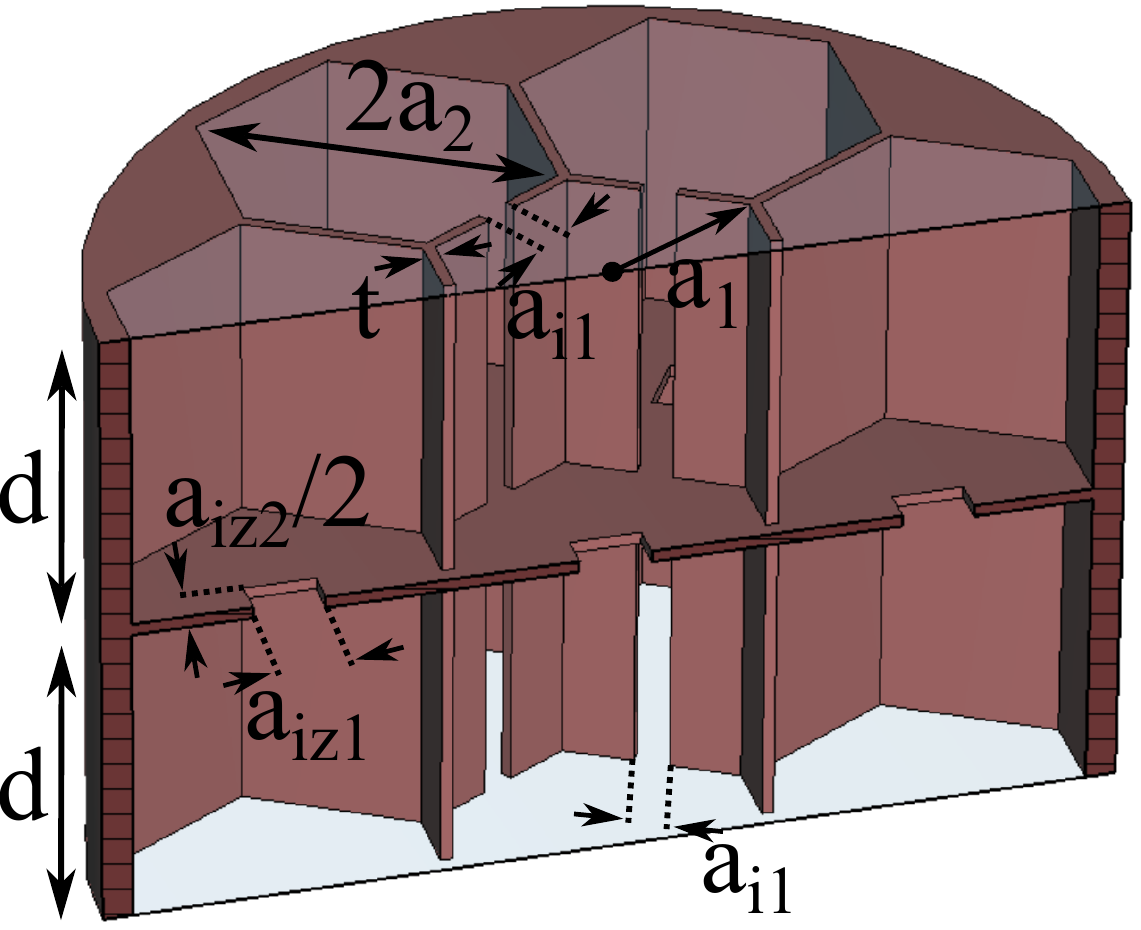}
         \caption{}
         \label{fig:HexCav_3D_14subcavities}
\end{subfigure}
\caption{3D models of hexagonal haloscopes for solenoid magnets: (a) single cavity, (b) 2D multicavity of 7 subcavities, and (c) 3D multicavity of 14 subcavities. In (b) and (c), light blue solids correspond with resonant cavities and brown solids with metallic iris sections. In (c), the symmetrical half of the structures is depicted. A thickness value of $t=1.5$~mm has been employed for all the scenarios.}
\label{fig:HexCav_SingleCavity_2Dmulticavity_and_3Dmulticavity}
\end{figure}

As previously stated, the conceptualization of 2D and 3D configurations is practically attainable solely for the $TM_{010}$ mode, primarily within solenoid magnet scenarios. Hence, the scope of this investigation pertains exclusively to this resonant mode for the designated structures. Within this paper, preliminary prototypes for 2D and 3D multicavities employing hexagonally structured subcavities have been formulated. In this analysis, the initial phase involves the design and optimization of an individual cavity of hexagonal geometry operating at $8.5$~GHz. A 3D model of this cavity is illustrated in Figure~\ref{fig:HexCav}.
The ensuing results of this preliminary endeavor are documented in Table~\ref{tab:HexCav_SingleCavity_2Dmulticavity_and_3Dmulticavity}. The outcomes are indicative of similar $Q_0V^2C^2$ factor values when compared with the data presented in Table~\ref{tab:SingleLongQVC} for a single circular cavity (solenoid magnet case).\\

\begin{table}[h]
\scalebox{0.8}{
\begin{tabular}{|c|c|c|c|c|c|c|c|}
\hline
Case and magnet & $2a$ (mm) & $d$ (mm) & $a_{i}$ (mm) & $V$ (mL) & $Q_0$ & $C$ & $Q_0V^2C^2$ (L$^2$) \\ \hline\hline
Single cav. and solenoid & $30$ & $26.39$ & $-$ & $15.4$ & $7.13\times10^4$ & $0.682$ & $7.875$ \\ \hline
2D multicav. and solenoid & $2a_1=29.15$ & $26.39$ & $3.4$ & $108.1$ & $7.19\times10^4$ & $0.622$ & $324.82$ \\
 & $2a_2=29.95$ & & & & & & \\ \hline
3D multicav. and solenoid & $2a_1=29.15$ & $26.39$ & $a_i=3.4$ & $213.61$ & $7.22\times10^4$ & $0.625$ & $1284.3$ \\
 & $2a_2=29.95$ & & $a_{iz1}=6$ & & & & \\
 & & & $a_{iz2}=6$ & & & & \\ \hline
\end{tabular}
}
\centering
\caption{\label{tab:HexCav_SingleCavity_2Dmulticavity_and_3Dmulticavity} Comparison of the operational parameters of a hexagonal single cavity, and 2D and 3D multicavities in a solenoid magnet ($TM_{010}$ mode).}
\end{table}

Using these numerical values as a basis, a duplication of the hexagonal resonator has been undertaken to construct a two-dimensional multicavity configuration consisting of seven subcavities. These subcavities are interconnected to the central cavity using inductive irises that are symmetrically positioned on the sidewalls, as depicted in Figure~\ref{fig:HexCav_2D_7subcavities}. Applying fine refinement in simulation, the dimensions and outcomes summarized in Table~\ref{tab:HexCav_SingleCavity_2Dmulticavity_and_3Dmulticavity} have been achieved. The resulting resonant frequency for this configuration stands at $8.49$~GHz, accompanied by a mode clustering value of approximately $\Delta f = 13$~MHz. It can be verified that the achieved $Q_0V^2C^2$ factor has notably expanded when compared to the hexagonal single cavity scenario. In a comparative analysis with the 2D case documented in Table~\ref{tab:CylCavsStackedInPhi2cav_and_Rho2cav_Solenoid_Allind} (pertaining to stacking along the $\varphi$ and $\rho$ axes), the achieved volume may appear slightly lower, yet the quality factor have experienced a considerable increase. This increase in the quality factor suggests that the potential for enhanced $Q_0V^2C^2$ factor is plausible following optimization efforts increasing the form factor value.\\

Concluding this investigation, a further examination was conducted involving the duplication of the previously detailed structure along its length, leading to the creation of a 3D multicavity configuration. This 3D construct is comprised of 14 hexagonal subcavities, each interconnected through inductive irises. All of these irises adhere to an inductive configuration. In the case of length-wise coupling, a selection was made in favour of centrally positioned square-shaped irises, as visually represented in Figure~\ref{fig:HexCav_3D_14subcavities}. The outcomes yielded by several optimization iterations are presented in Table~\ref{tab:HexCav_SingleCavity_2Dmulticavity_and_3Dmulticavity}. Specifically, the resonance frequency corresponding to this design registers at $8.502$~GHz, and the accompanying mode clustering value amounts to approximately $\Delta f = 12$~MHz.\\

As evident from the data, the reached $Q_0V^2C^2$ factor for this configuration closely aligns with that of the 3D structure devised for cylindrical cavities, as outlined in Table~\ref{tab:CylCavsStackedInLength2cav_and_Phi2cav_and_Rho2cav_Solenoid_Allind}. Nevertheless, it is worth noting that the design and optimization process of this structure is significantly simpler than the cylindrical scenario, due to the number of variables to be optimised. Hence, it can be asserted that the adoption of hexagonal resonant cavities presents an alternative approach for fabricating high-volume haloscopes through the implementation of multicavity configurations suitable for integration within solenoid magnets.

\subsection{Tuning}
\label{ss:2D3Dmulticavities_Tuning}

The techniques for adjusting resonant frequencies discussed in sections~\ref{ss:SingleCavities_Tuning} and \ref{ss:1Dmulticavities_Tuning} for single cavities and 1D multicavities can be similarly applied to the context of cylindrical or hexagonal structures based on 2D or 3D multicavities. The ADMX group has introduced a hybrid tuning system concept in \cite{Bowring:2018} which involves employing dielectric films and mechanical adjustments to tune hexagonal 2D and 3D multicavities.\\

In this case, individual tuning can be realized at each cavity, following the procedure described in \cite{Jeong:2018}. Nevertheless, tuning procedure for a multicavity is much more complex, since only the multicavity resonance is monitored.

\section{Conclusions and prospects}
\label{s:Conclusions}

This research delves into the exploration of volume limitations in cylindrical haloscopes. The enhancement of this parameter yields improvements in the $Q_0V^2C^2$, a well-known figure of merit of cavity haloscopes, thereby ultimately enhancing the scanning rate. Various strategies for augmenting the volume are presented, with considerations for constraints such as mode clustering which refers to the separation between adjacent modes, and the behaviour of form and quality factors. The study encompasses comprehensive investigations involving single cavities, 1D multicavities, and 2D and 3D multicavities that achieve significant $Q_0V^2C^2$ factors. The feasibility of implementing these haloscopes with the most substantial dipole and solenoid magnets in the axion community is demonstrated. The practical implementations of several designs yield favourable outcomes in quality factor, form factor, and mode clustering, effectively showcasing the potential of these studies while also validating their efficacy.\\

Among the various types of single cavities explored, it has been determined that large cavities operating within solenoid magnets ($TM_{010}$ mode) exhibit the most favorable performance in terms of the $Q_0V^2C^2$ figure of merit. Additionally, the utilization of multicavities, despite their increased design complexity has been demonstrated to enhance this factor. However, it is noted that the quest for increased volume to cover a range of axion masses is restricted by the permissible number of mode crossings. In the context of 1D multicavities, the feasibility of stacking in $\varphi$ and $\rho$ orientations for dipole magnets is found to be impractical. Among the extensive designs of large 1D multicavities conducted for solenoids, the structures stacked in $\rho$ orientation manifest the most favourable outcomes in terms of the $Q_0V^2C^2$ factor. For 2D multicavities, the alternative with the best figure of merit results is the haloscope stacked in $\rho$ and length. The 3D structure analysed has given very positive results in terms of $Q_0V^2C^2$ factor, exceeding the values of the single, 2D and 3D cases. In addition, a study with a single cavity, a 2D multicavity and a 3D multicavity for hexagonal geometry has been carried out, providing similar results to the cylindrical structures, but with simpler design and optimisation processes. These methodologies are intended to provide a practical guide for experimental axion research groups embarking on the pursuit of volume limits in the development of haloscopes based on cylindrical or hexagonal cavities, positioned within both dipole and solenoid magnets. Nevertheless, these approaches and analyses hold relevance for any scenario where augmenting the volume of the device at a specific frequency constitutes an objective.\\

This analysis opens up a range of promising strategies depending on the nature and configuration of the experiment magnet. The methodologies outlined in this research provide avenues to maximise the potential of magnet spacing, with the overall goal of refining the sensitivity and scanning rate of axion detection. In this sense, it is highly recommended to embark on exploring the limits of cylindrical geometries, taking advantage of various concepts proposed in this study. Examples include the incorporation of alternating couplings in 1D, 2D and 3D multicavities, as well as the implementation of elongated subcavities in 2D and 3D multicavities. In addition, envisaged exploration for the generalization for $N$ subcavities and in-depth analysis of different tuning procedures for multicavities is expected.

\acknowledgments
This work was performed within the RADES group. We thank our colleagues for their support. It has been funded by MCIN/AEI/10.13039/501100011033/ and by ``ERDF A way of making Europe'', under grants PID2019-108122GB-C33 and PID2022-137268NB-C53. JMGB thanks the grant FPI BES-2017-079787, funded by MCIN/AEI/10.13039/501100011033 and by "ESF Investing in your future", the European Research Council grant ERC-2018-StG-802836 (AxScale project), and the Lise Meitner program "In search of a new, light physics" of the Max Planck society. This article is based upon work from COST Action COSMIC WISPers CA21106, supported by COST (European Cooperation in Science and Technology).

\bibliographystyle{JHEP.bst}
\bibliography{mybibfile.bib}
\end{document}